\documentclass[12pt]{article}

\usepackage{epsf,amsfonts,hyperref}
\usepackage{graphicx}

\renewcommand{\appendix}[1]{
    \addtocounter{section}{1}
    \setcounter{equation}{0}
    \renewcommand{\thesection}{\Alph{section}}
    \section*{Appendix \thesection\protect\indent #1}
    \addcontentsline{toc}{section}{Appendix \thesection\ \ \ #1}
}
\newcommand\encadremath[1]{\vbox{\hrule\hbox{\vrule\kern8pt
\vbox{\kern8pt \hbox{$\displaystyle #1$}\kern8pt}
\kern8pt\vrule}\hrule}}
\def\enca#1{\vbox{\hrule\hbox{
\vrule\kern8pt\vbox{\kern8pt \hbox{$\displaystyle #1$}
\kern8pt} \kern8pt\vrule}\hrule}}

\newcommand\figureframex[3]{
\begin{figure}[bth]
\hrule\hbox{\vrule\kern8pt
\vbox{\kern8pt \vbox{
\begin{center}
{\mbox{\epsfxsize=#1.truecm\epsfbox{#2}}}
\end{center}
\caption{#3}
}\kern8pt}
\kern8pt\vrule}\hrule
\end{figure}
}
\newcommand\figureframey[3]{
\begin{figure}[bth]
\hrule\hbox{\vrule\kern8pt
\vbox{\kern8pt \vbox{
\begin{center}
{\mbox{\epsfysize=#1.truecm\epsfbox{#2}}}
\end{center}
\caption{#3}
}\kern8pt}
\kern8pt\vrule}\hrule
\end{figure}
}

\renewcommand{\thesection}{\arabic{section}}

\makeatletter
\@addtoreset{equation}{section}
\makeatother
\newtheorem{theorem}{Theorem}[section]

\newtheorem{remark}{Remark}[section]
\newtheorem{proposition}{Proposition}[section]
\newtheorem{lemma}{Lemma}[section]
\newtheorem{corollary}{Corollary}[section]
\newtheorem{definition}{Definition}[section]
\def\br{\begin{remark}\rm\small}
\def\er{\end{remark}}
\def\bt{\begin{theorem}}
\def\et{\end{theorem}}
\def\bd{\begin{definition}}
\def\ed{\end{definition}}
\def\bp{\begin{proposition}}
\def\ep{\end{proposition}}
\def\bl{\begin{lemma}}
\def\el{\end{lemma}}
\def\bc{\begin{corollary}}
\def\ec{\end{corollary}}
\def\beaq{\begin{eqnarray}}
\def\eeaq{\end{eqnarray}}
\newcommand{\proof}[1]{{\noindent \bf proof:}\par
{#1} $\square$}

\newcommand{\eq}[1]{Eq.~(\ref{#1})}

\newcommand{\beq}{\begin{equation}}
\newcommand{\eeq}{\end{equation}}
\newcommand{\bea}{\begin{eqnarray}}
\newcommand{\eea}{\end{eqnarray}}

\newcommand{\vs}{\vspace{0.7cm}}
\renewcommand{\and}{{\qquad {\rm and} \qquad}}

\newcommand{\virg}{{\qquad , \qquad}}

 \newcommand{\Tr}{{\,\rm Tr}\:}
\newcommand{\tr}{{\,\rm tr}\:}

\newcommand{\Res}{\mathop{\,\rm Res\,}}

\newcommand{\td}[1]{{\tilde{#1}}}

\newcommand{\ee}[1]{{{\rm e}^{#1}}}

\newcommand{\C}{{\mathbf C}}

\newcommand{\Pint}{{\int\kern -1.em -\kern-.25em}}

\newcommand{\acycle}{{\cal A}}
\newcommand{\bcycle}{{\cal B}}

\newcommand{\curve}{{\cal E}}

\newcommand{\bfa}{{\mathbf{a}}}
\newcommand{\bfb}{{\mathbf{b}}}
\newcommand{\bfalpha}{{\mathbf{\alpha}}}

\begin{document}
\sloppy


\pagestyle{empty}
\hfill SPhT-T07/055
\addtolength{\baselineskip}{0.20\baselineskip}
\begin{center}
\vspace{26pt}
{\large \bf {Topological expansion of mixed correlations in the hermitian 2 Matrix Model and x-y symmetry of the $F_g$
algebraic invariants.}}
\newline
\vspace{26pt}

{\sl B.\ Eynard}\hspace*{0.05cm}\footnote{ E-mail: eynard@spht.saclay.cea.fr }, {\sl N.\ Orantin}\hspace*{0.05cm}\footnote{ E-mail: orantin@spht.saclay.cea.fr }\\
\vspace{6pt}
Service de Physique Th\'{e}orique de Saclay,\\
F-91191 Gif-sur-Yvette Cedex, France.\\
\end{center}

\vspace{20pt}
\begin{center}
{\bf Abstract}:
We compute expectation values of mixed traces containing both matrices in a two matrix model,
i.e. generating function for counting bicolored discrete surfaces with non uniform boundary conditions.
As an application, we prove the $x-y$ symmetry of \cite{EOFg}.
\end{center}

%





\vspace{26pt}
\pagestyle{plain}
\setcounter{page}{1}


\newpage


\section{Introduction}

Formal matrix integrals can be regarded as an efficient toy model to explore the link between algebraic geometry and integrable systems \cite{Kri, Marco2}.
The theory of quantum gravity \cite{ZJDFG, davidRMT, KazakovRMT} is based on the idea that matrix models provide a generating function to measure ``volumes'' of moduli spaces of Riemann surfaces, and random matrix models were introduced in the 80's \cite{BIPZ} as a discretized version of 2d quantum gravity, i.e. conformal field theory coupled to gravity.

The formal matrix integral is at the same time a tau-function of some integrable hierarchy \cite{ZJDFG}, and it has a 't Hoof topological expansion \cite{thooft, ZJDFG, ACKM}:
\beq
\ln{\int_{\rm formal} dM \ee{-N\Tr V(M)}} = \sum_{g=0}^\infty N^{2-2g} F^{(g)}
\eeq
which is related to algebraic geometry (see \cite{MarcoF, Marco2, KazMar, eynm2m}).

In a recent work \cite{eynloop1mat, ec1loopF, eyno, CEO, EOFg}, we have developped a method to compute the $F^{(g)}$'s for various formal hermitian matrix models (1-matrix model, 2-matrix model, matrix model with an external field, double scaling limits of 2-matrix model) out of the data of an algebraic equation (called the classical spectral curve):
\beq
\curve(x,y)=0
\virg
\curve = {\rm polynomial}.
\eeq
The construction of \cite{EOFg} extends beyond matrix models, and the $F^{(g)}$'s can be computed for any algebraic equation of the type $\curve(x,y)=0$.

However the construction of \cite{EOFg} assumes an embedding of the curve into $\C^2$, i.e. the choice of 2 meromorphic functions $x$ and $y$ on the curve.
It was claimed in \cite{EOFg} that $F^{(g)}$ is invariant under the exchange $x\leftrightarrow y$, and the proof was announced to be published separately.

This is what we do in the present paper, together with additional results.

\bigskip

{\bf Mixed correlations}

In order to prove this claim, we first explore the case where the $F^{(g)}$'s come from a formal 2-matrix model (the symmetry $x\leftrightarrow y$ holds almost by definition in that case, see \cite{CEO}). 
We write the loop equation relations (W-algebra) \cite{staudacher,eynchain}, which we solve,
and we are led to define new mixed correlation functions ($W_{k,l}$ and $H_{k,l}$ below), which did not appear in \cite{EOFg}.

In the application of the 2-matrix model to quantum gravity and conformal field theory, those mixed correlation functions were known to play an important role in the understanding of boundary operators.
But their explicit computation has been a challenge until recently. The main reason is that they don't reduce to eigenvalues of the matrices, and could not be computed by standard methods.
The first explicit computations were obtained in \cite{BEmixed} and \cite{eynprats}.
Here in this paper, we show how to compute the topological expansion of a family of mixed correlation functions of the 2-matrix model.
In a coming work \cite{EOallmixed}, we shall show how to compute all mixed correlations, and introduce a link with group theory and Bethe ansatz (this is a generalization of \cite{EObethe}).

\bigskip

Then, for the general case (i.e. if $\curve$ was not obtained from a matrix model), we mimic those mixed correlation functions and that allows to prove the $x\leftrightarrow y$ symmetry of $F^{(g)}$.

\section{Mixed traces of matrix models}

Consider the formal 2-matrix integral\footnote{A formal integral is defined as a formal power series in some expansion parameter $t$, as explained in \cite{eynform} or \cite{EOFg}. Formal matrix integrals always have a $1/N^2$ expansion order by order in $t$, called the topological expansion.}:
\beq
Z = \int dM_1\, dM_2\,\, \ee{-N\tr(V_1(M_1)+V_2(M_2)-M_1 M_2)}
\eeq
where we assume in this section that $V_1$ is a polynomial of degree $d_1+1$ and $V_2$ is a polynomial of degree $d_2+1$.

Our goal is to compute the following connected expectation values:
\bea
&& \overline{W}_{k,l}(x_1,\dots,x_k|y_1,\dots,y_l)  \cr
&=& \left<\tr{1\over x_1-M_1}\tr{1\over x_2-M_1}\dots\tr{1\over x_k-M_1}\,\,\,\tr{1\over y_1-M_2}\tr{1\over y_2-M_2}\dots\tr{1\over y_l-M_2}\right>_c \cr
&=& \sum_{g=0}^\infty N^{2-2g-k-l}\,\overline{W}_{k,l}^{(g)}(x_1,\dots,x_k|y_1,\dots,y_l).
\eea
and
\bea
&& {\overline{H}}_{k,l}(x,y;x_1,\dots,x_k|y_1,\dots,y_l) \cr
&=& \left<\tr{1\over x-M_1}{1\over y-M_2}\,\, \tr{1\over x_1-M_1}\dots\tr{1\over x_k-M_1}\,\,\,\tr{1\over y_1-M_2}\dots\tr{1\over y_l-M_2}\right>_c  \cr
&=& \sum_{g=0}^\infty N^{2-2g-k-l-1}\,\overline{H}_{k,l}^{(g)}(x,y;x_1,\dots,x_k|y_1,\dots,y_l)
\eea
$\overline{W}_{k,l}^{(g)}$ is the generating function which counts connected genus $g$ bi-colored discrete surfaces with $k$ boundaries of the first color, and $l$ boundaries of the second color.
$\overline{H}_{k,l}^{(g)}$ is the generating function which counts genus $g$ bi-colored discrete surfaces with $k$ boundaries of the first color, and $l$ boundaries of the second color, and one additional boundary which carries the 2 colors.
The power of $N$ in both cases is the Euler characteristic of such surfaces.
The 2-matrix model was introduced in \cite{KazakovIsing} as a discrete version of the Ising model on a random surface.

Notice that in $\overline{H}_{k,l}^{(g)}$, the first trace contains both matrices $M_1$ and $M_2$, we call it a {\bf mixed trace} because it cannot be expressed in terms of eigenvalues of $M_1$ and $M_2$.
In applications of matrix models to conformal field theories, such objects correspond to the insertion of a pair of boundary operators, and are thus very interesting.
$\overline{H}_{0,0}^{(0)}$ was computed in many works \cite{eynchain, DKK}, and in the context of convergent integrals (instead of formal integrals), $\overline{H}_{0,0}$ was computed in \cite{BEmixed,eynprats, BergereEyn}.

\bigskip

The $\overline{W}_{k,0}^{(g)}$'s were already computed in \cite{eynloop1mat, eyno, CEO}, and are given by the algebraic invariants defined in \cite{EOFg}, they are the non mixed traces.

\medskip

It is known (see for instance \cite{CEO}) that all those functions are multivalued functions of their $x$ or $y$ variables, and they are in fact functions living on a Riemann surface called the spectral curve of equation:
\beq
\curve(x,y)=0.
\eeq
On this curve, we chose a canonical basis of cycles\footnote{All required definitions relative to algebraic geometry can be found in \cite{EOFg} or more generally
in \cite{Fay,Farkas}. We will use all along these notes the notations of \cite{EOFg}.
The ${\cal{A}}$ and ${\cal{B}}$-cycles may be the modified cycles of \cite{EOFg}.} ${\cal{A}}_i \cap {\cal{B}}_j = \delta_{i,j}$, $i,j=1,\dots{\cal{G}}$, where ${\cal{G}}$ denotes the genus of the curve $\curve$.
We will note by $p^i$ (resp. $\tilde{p}^j$) the different points of $\curve$ whose projection in the complex plane by the meromorphic
function $x$ (resp. $y$) are equal:
\beq
\forall i = 1 \dots d_2 \, , \; x(p^i) = x(p^0)
\virg
\forall i = 1 \dots d_1 \, , \; y(\tilde{p}^i) = x(\tilde{p}^0),
\eeq
where the superscript $0$ refers to the $x$- and $y$-physical sheets.

It is thus more convenient to redefine $\overline{W}_{k,l}^{(g)}$ and $\overline{H}_{k,l}^{(g)}$ in terms of meromorphic forms on the curve:
\bea
&& W_{k,l}^{(g)}(p_1,\dots,p_k|q_1,\dots,q_l)  \cr
&=& \overline{W}_{k,l}^{(g)}(x(p_1),\dots,x(p_k)|y(q_1),\dots,y(q_l))\,\, dx(p_1)\dots dx(p_k) dy(q_1)\dots  dy(q_l) \cr
&& + \delta_{g,0}\delta_{k,1}\delta_{l,0} (y(p_1)-V'_1(x(p_1))) dx(p_1) + \delta_{g,0}\delta_{k,0}\delta_{l,1} (x(q_1)-V'_2(y(q_1))) dy(q_1) \cr
&& + {\delta_{g,0}\delta_{k,2}\delta_{l,0}\,\, dx(p_1)dx(p_2) \over (x(p_1)-x(p_2))^2 }  \cr
&& + {\delta_{g,0}\delta_{k,0}\delta_{l,2}\,\, dy(q_1)dy(q_2) \over (y(q_1)-y(q_2))^2 }  \cr
\eea
where the $p_i$'s and $q_j$'s are now points on the curve $\curve$, instead of points in the complex plane.
We have also "renormalized the unstable functions'' with $2-2g-k-l \geq 0$.

With those notations we have \cite{CEO, MarcoF}:
\beq
W_{1,0}^{(0)}=W_{0,1}^{(0)} = 0,
\eeq
\beq
W_{2,0}^{(0)}(p,q) =-W_{1,1}^{(0)}(p,q)=W_{0,2}^{(0)}(p,q) = B(p,q)
\eeq
where $B$ is the Bergmann kernel, i.e. the unique bilinear form on $\curve$ with a double pole at $p=q$ and no other pole, with vanishing residue, and normalized on $\acycle$-cycles:
\beq
B(p,q) \mathop{\sim}_{p\to q} {dz(p)dz(q) \over (z(p)-z(q))^2 } + {\rm finite}
\virg
\forall i = 1 \dots {\cal{G}} \, , \;
\oint_{\acycle} B =0.
\eeq

We also define the differentials corresponding to the mixed correlation functions:
\bea
&& H_{k,l}^{(g)}(p,q;p_1,\dots,p_k|q_1,\dots,q_l)  \cr
&=& \overline{H}_{k,l}^{(g)}(x(p),y(q);x(p_1),\dots,x(p_k)|y(q_1),\dots,y(q_l))\,\, dx(p_1)\dots dx(p_k) dy(q_1)\dots  dy(q_l) \cr
&& + \delta_{g,0}\delta_{k,0}\delta_{l,0}
\eea
and we normalize them by the leading order of the simplest mixed correlation function:
\bea
h_{k,l}^{(g)}(p,q;p_1,\dots,p_k|q_1,\dots,q_l)
= {H_{k,l}^{(g)}(p,q;p_1,\dots,p_k|q_1,\dots,q_l)  \over H_{0,0}^{(0)}(p,q)  }.
\eea

It is well known \cite{eynm2m, eynchain, DKK} (and it can be rederived from \eq{loopeqUH} and \eq{loopeqEU} below) that:
\beq
H_{0,0}^{(0)}(p,q) = {\curve(x(p),y(q))\over (x(p)-x(q))(y(p)-y(q))}.
\eeq

We also need to introduce:
\bea
&& U_{k,l}(p,y;p_1,\dots,p_k|q_1,\dots,q_l) \cr
&=& \Big<\tr{1\over x(p)-M_1}{V'_2(y)-V'_2(M_2)\over y-M_2}\,\, \tr{dx(p_1)\over x(p_1)-M_1}\dots\tr{dx(p_k)\over x(p_k)-M_1}\,\,\,\cr
&& \tr{dy(q_1)\over y(q_1)-M_2}\dots\tr{dy(q_l)\over y(q_l)-M_2}\Big>_c  \cr
&& + \delta_{g,0}\delta_{k,0}\delta_{l,0} (V'_2(y)-x(p)) \cr
&=& \sum_{g=0}^\infty N^{2-2g-k-l-1}\,U_{k,l}^{(g)}(p,y;p_1,\dots,p_k|q_1,\dots,q_l),
\eea
which is a polynomial of $y$ of degree at most $d_2-1$,
\bea
&& \td{U}_{k,l}(x,q;p_1,\dots,p_k|q_1,\dots,q_l) \cr
&=& \Big<\tr{V'_1(x)-V'_1(M_1)\over x-M_1}{1\over y(q)-M_2}\,\, \tr{dx(p_1)\over x(p_1)-M_1}\dots\tr{dx(p_k)\over x(p_k)-M_1}\,\,\,\cr
&& \tr{dy(q_1)\over y(q_1)-M_2}\dots\tr{dy(q_l)\over y(q_l)-M_2}\Big>_c  \cr
&& + \delta_{g,0}\delta_{k,0}\delta_{l,0} (V'_1(x)-y(p)) \cr
&=& \sum_{g=0}^\infty N^{2-2g-k-l-1}\,\td{U}_{k,l}^{(g)}(x,q;p_1,\dots,p_k|q_1,\dots,q_l),
\eea
which is a polynomial of $x$ of degree at most $d_1-1$ and
\bea
&& -E_{k,l}(x,y;p_1,\dots,p_k|q_1,\dots,q_l) \cr
&=& \Big<\tr{V'_1(x)-V'_1(M_1)\over x-M_1}{V'_2(y)-V'_2(M_2)\over y-M_2}\,\, \tr{dx(p_1)\over x(p_1)-M_1}\dots\tr{dx(p_k)\over x(p_k)-M_1}\,\,\,\cr
&& \tr{dy(q_1)\over y(q_1)-M_2}\dots\tr{dy(q_l)\over y(q_l)-M_2}\Big>_c  \cr
&& + \delta_{g,0}\delta_{k,0}\delta_{l,0} ((V'_1(x)-y(p))(V'_2(y)-x(p)) -1) \cr
&=& -\sum_{g=0}^\infty N^{2-2g-k-l-1}\,E_{k,l}^{(g)}(x,y;p_1,\dots,p_k|q_1,\dots,q_l),
\eea
which is a polynomial of $x$ of degree $d_1-1$ and of $y$ of degree $d_2-1$.

We have:
\beq
E_{0,0}^{(0)}(x,y) = \curve(x,y)
\virg
U_{0,0}^{(0)}(p,y) = {\curve(x(p),y)\over y-y(p)}
\virg
\td{U}_{0,0}^{(0)}(x,q) = {\curve(x,y(q))\over x-x(q)},
\eeq
and
\beq
P_{0,0}^{(0)}(x,y) = - \curve(x,y).
\eeq

\subsection{Loop equations}

In order to obtain a closed set of equations computing these mixed correlation functions, we consider 4 families of loop
equations \cite{staudacher, eynchain, eylooprat} corresponding to different infinitesimal changes of variables $M_i \to M_i + \epsilon\delta M_i$ in the matrix integral.

$\delta M_2 = {1 \over x(p)-M_1} {1 \over y(q)-M_2} {\displaystyle \prod_{i=1}^k} \tr {1 \over x(p_i)-M_1} {\displaystyle \prod_{j=1}^l} \tr {1 \over y(q_j)-M_2}$
gives:
\bea\label{loopeqUH}
 - U^{(g)}_{k,l}(p,y(q);{\bf p_K}|{\bf q_L})
&=& (x(p)-x(q))H^{(g)}_{k,l}(p,q;{\bf p_K}|{\bf q_L}) \cr
&& + \sum_h\sum_{I,J} {W^{(h)}_{i,j+1}({\bf p_I} |{\bf q_J},q)  H^{(g-h)}_{k-i,l-j}(p,q;{\bf p_{K/I}}|{\bf q_{L/J}})\over dy(q)} \cr
&& +  {H^{(g-1)}_{k,l+1}(p,q;{\bf p_K}|{\bf q_L},q)\over dy(q)} \cr
&& - \sum_n d_{q_n}\,{H^{(g)}_{k,l-1}(p,q_n;{\bf p_{K}}|{\bf q_{L/\{n\}}})\over y(q)-y(q_n)}
\eea
$\delta M_1 = {1 \over x(p)-M_1} {1 \over y(q)-M_2} {\displaystyle \prod_{i=1}^k} \tr {1 \over x(p_i)-M_1} {\displaystyle \prod_{j=1}^l} \tr {1 \over y(q_j)-M_2}$
gives:
\bea\label{loopeqtdUH}
 - \td{U}^{(g)}_{k,l}(x(p),q;{\bf p_K}|{\bf q_L})
&=& (y(q)-y(p))H^{(g)}_{k,l}(p,q;{\bf p_K}|{\bf q_L}) \cr
&& + \sum_h\sum_{I,J} {W^{(h)}_{i+1,j}(p,{\bf p_I}|{\bf q_J})  H^{(g-h)}_{k-i,l-j}(p,q;{\bf p_{K/I}}|{\bf q_{L/J}})\over dx(p)} \cr
&& +  {H^{(g-1)}_{k+1,l}(p,q;p,{\bf p_K}|{\bf q_L})\over dx(p)} \cr
&& - \sum_m d_{p_m}\,{H^{(g)}_{k-1,l}(p_m,q;{\bf p_{K/\{m\} }}|{\bf q_L})\over x(p)-x(p_m)}
\eea
$\delta M_2 = {V_1'(x(p))-V_1'(M_1) \over x(p)-M_1} {1 \over y(q)-M_2} {\displaystyle \prod_{i=1}^k} \tr {1 \over x(p_i)-M_1} {\displaystyle \prod_{j=1}^l} \tr {1 \over y(q_j)-M_2}$
gives:
\bea\label{loopeqEtdU}
 E^{(g)}_{k,l}(x(p),y(q);{\bf p_K}|{\bf q_L})
&=& (x(p)-x(q)) \td{U}^{(g)}_{k,l}(x(p),q;{\bf p_K}|{\bf q_L}) \cr
&& + \sum_h \sum_{I,J} {\check{W}^{(h)}_{i,j+1}({\bf p_I}|{\bf q_J},q)  \td{U}^{(g-h)}_{k-i,l-j}(x(p),q;{\bf p_{K/I}}|{\bf q_{L/J}})\over dy(q)} \cr
&& + {\td{U}^{(g-1)}_{k,l+1}(x(p),q;{\bf p_K}|{\bf q_L},q)\over dy(q)} \cr
&& - \sum_m d_{q_m}\,{\td{U}^{(g)}_{k,l-1}(x(p),q_m;{\bf p_K}|{\bf q_{L/\{m\}}})\over y(q)-y(q_m)} \cr
&& - \sum_m d_{p_m}\,H^{(g)}_{k-1,l}(p_m,q;{\bf p_{K/\{m\} }}|{\bf q_L})
\eea
and $\delta M_1 = {1 \over x(p)-M_1} {V_2'(y(q))-V_2'(M_2) \over y(q)-M_2} {\displaystyle \prod_{i=1}^k} \tr {1 \over x(p_i)-M_1} {\displaystyle \prod_{j=1}^l} \tr {1 \over y(q_j)-M_2}$
gives:
\bea\label{loopeqEU}
 E^{(g)}_{k,l}(x(p),y(q);{\bf p_K}|{\bf q_L})
&=& (y(q)-y(p)) U^{(g)}_{k,l}(p,y(q);{\bf p_K}|{\bf q_L}) \cr
&& + \sum_h \sum_{I,J} {{W}^{(h)}_{i+1,j}(p,{\bf p_I}|{\bf q_J})  U^{(g-h)}_{k-i,l-j}(p,y(q);{\bf p_{K/I}}|{\bf q_{L/J}})\over dx(p)} \cr
&& + {U^{(g-1)}_{k+1,l}(p,y(q);p,{\bf p_K}|{\bf q_L})\over dx(p)} \cr
&& - \sum_m d_{p_m}\,{U^{(g)}_{k-1,l}(p_m,y(q);{\bf p_{K/\{m\}}}|{\bf q_L}) \over x(p)-x(p_m)} \cr
&& - \sum_m d_{q_m}\,H^{(g)}_{k,l-1}(p,q_m;{\bf p_K}|{\bf q_{L/\{m\}}}).
\eea

Those loop equations can be seen to be equivalent to W-algebra constraints  \cite{Virasoro, ZJDFG}, or to a generalization of Tutte's equations for the combinatorics of discrete surfaces \cite{tutte,tutte2}.

\subsection{Solution of loop equations}

\bt\label{thmixedtraces}
The solution of loop equations is such that:

\bea\label{defrechkl}
&& h_{k,l}^{(g)}(p,q;{\bf p_K}|{\bf q_L}) \cr
&=& \Res_{r\to \td{q}^j,p,{\bf p_K}} {1\over (x(p)-x(r))(y(r)-y(q))}\,\Big( h_{k+1,l}^{(g-1)}(r,q;r,{\bf p_{K}}|{\bf q_{L}})   \cr
&& + \sum_h \sum_{I\subset K}\sum_{J\subset L} W_{i+1,j}^{(h)}(r,{\bf p_I}|{\bf q_J}) h_{k-i,l-j}^{(g-h)}(r,q;{\bf p_{K/I}}|{\bf q_{L/J}}) \Big), \cr
\eea
\bea\label{defrecWkl}
&& W_{k,l+1}^{(g)}({\bf p_K}|{\bf q_L},q) \cr
&=& \Res_{r\to \td{q}^j,{\bf p_K}} {dy(q)\over (y(r)-y(q))}\,\Big( h_{k+1,l}^{(g-1)}(r,q;r,{\bf p_{K}}|{\bf q_{L}}) \cr
&& + \sum_h \sum_{I\subset K}\sum_{J\subset L} W_{i+1,j}^{(h)}(r,{\bf p_I}|{\bf q_J}) h_{k-i,l-j}^{(g-h)}(r,q;{\bf p_{K/I}}|{\bf q_{L/J}}) \Big). \cr
\eea
where ${\displaystyle \Res_{r \to \td{q}^j}}$ means that one takes the residues around all the points $\td{q}^j \neq q$ such that
$y(\td{q}^j) = y(q)$.
\et

Given the initial conditions:
\beq
h^{(0)}_{0,0}=1
\virg
W^{(g)}_{k,0}(p_1,\dots,p_k)= \left.{W}^{(g)}_{k}(p_1,\dots,p_k)\right|_{\curve}
\eeq
where $\left.{W}^{(g)}_{k}(p_1,\dots,p_k)\right|_{\curve}$ is the function defined in \cite{EOFg},
the above system is triangular and computes univocally any  $h^{(g)}_{k,l}$ and $W^{(g)}_{k,l}$ in at most $k+l+{g^2\over 2}$ steps.

One easily proves by recursion on $2g+k+l$ that:
\beq
{H}_{k,l}^{(g)}(p,q;{\bf p_K}|{\bf q_L})\,\,\, {\rm has\, poles}\,
\left\{
\begin{array}{l}
{\rm in}\,\, p=a,q,{\bf q_L} \cr
{\rm in}\,\, q=b,p,{\bf p_K} \cr
{\rm in}\,\, p_j =a,q,{\bf q_L} \cr
{\rm in}\,\, q_j =b,p,{\bf p_K} \cr
\end{array}
\right.
\eeq
and
\beq
{W}_{k,l}^{(g)}({\bf p_K}|{\bf q_L})\,\,\, {\rm has\, poles}\,
\left\{
\begin{array}{l}
{\rm in}\,\, p_j =a,{\bf q_L} \cr
{\rm in}\,\, q_j =b,{\bf p_K} \cr
\end{array}
\right.
\eeq

\proof{
Since $\td{U}^{(g)}_{k,l}(x(p),q;{\bf p_K}|{\bf q_L})$ is a polynomial in $x(p)$ of degree at most $d_1-2$, it is given by the Lagrange interpolation formula:
\bea
\td{U}^{(g)}_{k,l}(x(p),q;{\bf p_K}|{\bf q_L})
&=& \td{U}^{(0)}_{0,0}(x(p),q) { \sum_{j=1}^{d_1}} {\td{U}^{(g)}_{k,l}(x(\td{q}^j),q;{\bf p_K}|{\bf q_L}) \over (x(p)-x(\td{q}^j)) \td{U}^{(0)}_{0,0\,\, x}(x(\td{q}^j),q)}  \cr
&=& \td{U}^{(0)}_{0,0}(x(p),q) \sum_{j=1}^{d_1} \Res_{r\to \td{q}^j} {\td{U}^{(g)}_{k,l}(x(\td{q}^j),q;{\bf p_K}|{\bf q_L}) \, dx(r)\over (x(p)-x(r))\,\td{U}^{(0)}_{0,0}(x(r),q)} . \cr
\eea
Then we replace $\td{U}^{(g)}_{k,l}(x(\td{q}^j),q;{\bf p_K}|{\bf q_L})$ by its value from the loop equation \ref{loopeqtdUH}:
\beq
\begin{array}{rcl}
\td{U}^{(g)}_{k,l}(x(p),q;{\bf p_K}|{\bf q_L})
&=& -  {\displaystyle \sum_{j=1}^{d_1}} \Res_{r\to \td{q}^j} {\td{U}^{(0)}_{0,0}(x(p),q)\over (x(p)-x(r))\,\td{U}^{(0)}_{0,0}(x(r),q)} \left[\phantom{{H^{(g)}_{k-1,l}(p_m,q;p_{K/\{m\} }|q_L)\,dx(r)\over x(r)-x(p_m)}} \right.\cr
&& {\displaystyle \sum_h\sum_{I,J}} {W^{(h)}_{i+1,j}(r,{\bf p_I} |{\bf q_J})  H^{(g-h)}_{k-i,l-j}(r,q;{\bf p_{K/I}}|{\bf q_{L/J}})} \cr
&& \left. +  {H^{(g-1)}_{k+1,l}(r,q;r,{\bf p_K}|{\bf q_L})}  - {\displaystyle \sum_m} d_{p_m}\,{H^{(g)}_{k-1,l}(p_m,q;{\bf p_{K/\{m\} }}|{\bf q_L})\,dx(r)\over x(r)-x(p_m)}
\right]  \cr
\end{array}
\eeq

Notice that the same residue computed at $r\to p$ gives the terms in the RHS of the loop equation \ref{loopeqtdUH}, and therefore:
\bea
&& (y(q)-y(p)) H^{(g)}_{k,l}(p,q;{\bf p_K}|{\bf q_L}) \cr
&=&   \Res_{r\to p,\td{q}^j} {\td{U}^{(0)}_{0,0}(x(p),q)\over (x(p)-x(r))\,\td{U}^{(0)}_{0,0}(x(r),q)} \left[\phantom{{H^{(g)}_{k-1,l}(p_m,q;p_{K/\{m\} }|q_L)\,dx(r)\over x(r)-x(p_m)}} \right.\cr
&&  \sum_h\sum_{I,J} {W^{(h)}_{i+1,j}(r,{\bf p_I} |{\bf q_J})  H^{(g-h)}_{k-i,l-j}(r,q;{\bf p_{K/I}}|{\bf q_{L/J}})} \cr
&& \left. +  {H^{(g-1)}_{k+1,l}(r,q;r,{\bf p_K}|{\bf q_L})}  - \sum_m d_{p_m}\,{H^{(g)}_{k-1,l}(p_m,q;{\bf p_{K/\{m\} }}|{\bf q_L})\,dx(r)\over x(r)-x(p_m)}
\right].  \cr
\eea
Moreover the last term $d_{p_m}\,{H^{(g)}_{k-1,l}(p_m,q;{\bf p_{K/\{m\} }}|{\bf q_L})\,dx(r)\over x(r)-x(p_m)}$ can be computed explicitely:
\beq
\begin{array}{l}
  d_{p_m}  \Res_{r\to p,\td{q}^j} {\td{U}^{(0)}_{0,0}(x(p),q)\over (x(p)-x(r))\,\td{U}^{(0)}_{0,0}(x(r),q)}\, {H^{(g)}_{k-1,l}(p_m,q;{\bf p_{K/\{m\} }}|{\bf q_L})\,dx(r)\over x(r)-x(p_m)} \cr
=  d_{p_m}  \Res_{r\to p,\td{q}^j} {\curve(x(p),y(q))(x(r)-x(q))\over (x(p)-x(r))(x(p)-x(q))\,\curve(x(r),y(q))} \,{H^{(g)}_{k-1,l}(p_m,q;{\bf p_{K/\{m\} }}|{\bf q_L})\,dx(r)\over x(r)-x(p_m)}. \cr
\end{array}
\eeq
Under this form, one can see that the integrant is a rational function of $x(r)$. Thus, the residue can be computed on
the complex plane obtained by the projection $x$ and we can move the integration contours on the complex plane instead of
the curve $\curve$ itself. This term is then equal to:
\bea
&& d_{p_m}  \Res_{x\to x(p),x(\td{q}^j)} {\curve(x(p),y(q))(x-x(q))\over (x(p)-x)(x(p)-x(q))\,\curve(x,y(q))} \,{H^{(g)}_{k-1,l}(p_m,q;{\bf p_{K/\{m\} }}|{\bf q_L})\,dx\over x-x(p_m)} \cr
&=& - d_{p_m}  \Res_{x\to x(p_m)} {\curve(x(p),y(q))(x-x(q))\over (x(p)-x)(x(p)-x(q))\,\curve(x,y(q))} \,{H^{(g)}_{k-1,l}(p_m,q;{\bf p_{K/\{m\} }}|{\bf q_L})\,dx\over x-x(p_m)} \cr
&=& - d_{p_m}   {\curve(x(p),y(q))(x(p_m)-x(q))\over (x(p)-x(p_m))(x(p)-x(q))\,\curve(x(p_m),y(q))} \,H^{(g)}_{k-1,l}(p_m,q;{\bf p_{K/\{m\} }}|{\bf q_L}) \cr
&=& - \Res_{r\to p_m}   {\td{U}^{(0)}_{0,0}(x(p),q)\over (x(p)-x(r))\,\td{U}^{(0)}_{0,0}(x(r),q)} \,H^{(g)}_{k-1,l}(r,q;{\bf p_{K/\{m\} }}|{\bf q_L}) W_{2,0}^{(0)}(r,p_m) \cr
&=& - \Res_{r\to p_m}   {\td{U}^{(0)}_{0,0}(x(p),q)\over (x(p)-x(r))\,\td{U}^{(0)}_{0,0}(x(r),q)}
\Big(  \sum_{h,I,J} {W^{(h)}_{i+1,j}(r,{\bf p_I} |{\bf q_J})  H^{(g-h)}_{k-i,l-j}(r,q;{\bf p_{K/I}}|{\bf q_{L/J}})} \cr
&& +  {H^{(g-1)}_{k+1,l}(r,q;r,{\bf p_K}|{\bf q_L})} \Big),
\eea
where the last equality holds thanks to the loop equation \eq{loopeqtdUH}.
Therefore:
\bea
 (y(q)-y(p)) H^{(g)}_{k,l}(p,q;{\bf p_K}|{\bf q_L})
&=&   \Res_{r\to p,\td{q}^j,{\bf p_K}} {\td{U}^{(0)}_{0,0}(x(p),q)\over (x(p)-x(r))\,\td{U}^{(0)}_{0,0}(x(r),q)} \Big( \cr
&&  \sum_h\sum_{I,J} {W^{(h)}_{i+1,j}(r,{\bf p_I} |{\bf q_J})  H^{(g-h)}_{k-i,l-j}(r,q;{\bf p_{K/I}}|{\bf q_{L/J}})} \cr
&& +  {H^{(g-1)}_{k+1,l}(r,q;r,{\bf p_K}|{\bf q_L})}  \Big).
\eea

If we divide by $\td{U}^{(0)}_{0,0}(x(p),q)$ we obtain:
\bea
 - h^{(g)}_{k,l}(p,q;{\bf p_K}|{\bf q_L})
&=&   \Res_{r\to p,\td{q}^j,p_K} {1\over (x(p)-x(r))\,(y(r)-y(q))} \Big( \cr
&&  \sum_h\sum_{I,J} {W^{(h)}_{i+1,j}(r,{\bf p_I} |{\bf q_J})  h^{(g-h)}_{k-i,l-j}(r,q;{\bf p_{K/I}}|{\bf q_{L/J}})} \cr
&& +  {h^{(g-1)}_{k+1,l}(r,q;r,{\bf p_K}|{\bf q_L})}  \Big).
\eea

The other half of the theorem is obtained from the fact that for large $x$:
\beq
\tr {1\over x-M_1}{1\over y-M_2} \to {1\over x} \tr {1\over y-M_2}
\eeq
and thus:
\beq
H^{(g)}_{k,l}(p,q;{\bf p_K}|{\bf q_L})
\to {1\over x(p)}
{W^{(g)}_{k,l+1}({\bf p_K}|{\bf q_L},q)\over dy(q)}
\eeq
when $p \to \infty_x$\footnote{$\infty_x$ is the only point on the curve where the meromorphic function $x$ has a simple pole (see \cite{eynm2mg1} for further details).}.
}

\subsection{Examples, first few terms}

Let us solve the recursive definition and give explicit formulae for the simplest functions.

\medskip
{\bf Example $W_{1,1}^{(0)}$:}

In particular, definitions \eq{defrechkl} and \eq{defrecWkl} give:
\bea
W_{1,1}^{(0)}(p_1|q)
&=& \Res_{r\to \td{q}^j,p_1} {dy(q)\,B(r,p_1)\over (y(r)-y(q))}   \cr
&=& - \Res_{r\to q} {dy(q)\,B(r,p_1)\over (y(r)-y(q))}   \cr
&=& - B(q,p_1).
\eea
Therefore we recover:
\beq\label{A100}
W_{2,0}^{(0)}(p_1,q)+W_{1,1}^{(0)}(p_1|q)=0.
\eeq

\medskip
{\bf Example $H_{1,0}^{(0)}$:}

\bea
h_{1,0}^{(0)}(p,q;p_1)
&=& \Res_{r\to \td{q}^j,p,p_1} {B(r,p_1)\over (x(p)-x(r))(y(r)-y(q))} \cr
&=& -\Res_{r\to p^i,q} {B(r,p_1)\over (x(p)-x(r))(y(r)-y(q))}.
\eea

\medskip
{\bf Example $H_{0,1}^{(0)}$:}

\bea
h_{0,1}^{(0)}(p,q;p_1)
&=& \Res_{r\to \td{q}^j,p} { W_{1,1}^{(0)}(r|p_1)\over (x(p)-x(r))(y(r)-y(q))}  \cr
&=& - \Res_{r\to \td{q}^j,p} { B(r,p_1)\over (x(p)-x(r))(y(r)-y(q))}  \cr
&=& \Res_{r\to p^i,q,p_1} { B(r,p_1)\over (x(p)-x(r))(y(r)-y(q))}.
\eea

Moreover we have:
\bea\label{B000}
h_{1,0}^{(0)}(p,q;p_1)+h_{0,1}^{(0)}(p,q;p_1)
&=& \Res_{r\to p_1} {B(r,p_1)\over (x(p)-x(r))(y(r)-y(q))} \cr
&=& d_{p_1} \left( {1\over (x(p)-x(p_1))(y(p_1)-y(q))} \right).
\eea

\medskip
{\bf Example $W_{2,1}^{(0)}$:}

\beq
\begin{array}{rcl}
{W_{2,1}^{(0)}(p_1,p_2|q)\over dy(q)}
&=&  \Res_{r\to \td{q}^j,p_1,p_2} {B(r,p_1)h_{1,0}^{(0)}(r,q;p_2)+B(r,p_2)h_{1,0}^{(0)}(r,q;p_1)+W_{3,0}^{(0)}(r,p_1,p_2)\over (y(r)-y(q))} \cr
&=& - \Res_{r\to q,\bfa} {B(r,p_1)h_{1,0}^{(0)}(r,q;p_2)+B(r,p_2)h_{1,0}^{(0)}(r,q;p_1)+W_{3,0}^{(0)}(r,p_1,p_2)\over (y(r)-y(q))}. \cr
\end{array}
\eeq

\subsection{Conclusion of section 2}

Therefore, through theorem \ref{thmixedtraces}, we have an effective explicit method to compute any
$H_{k,l}^{(g)}$ and any $W_{k,l}^{(g)}$ for the 2-matrix model.

This is an interesting result in itself, since none of those quantities were computed before, and
those quantities are of importance in applications of random matrices to combinatorics of maps with colored boundaries, i.e. boundary conformal field theory.

\bigskip

An important remark, is that we have chosen to emphasize the role of the loop equation \ref{loopeqtdUH}, rather than equation \ref{loopeqUH}, i.e. we have used the Lagrange interpolation formula for a polynomial in $x$, whereas we could have done the same thing with a polynomial in $y$.
In other words, we have chosen the $x$-representation rather than the $y$-representation, although both methods {\bf must} give the same answer.
In particular, given $W_{k,0}$, theorem \ref{thmixedtraces} allows to compute $W_{0,l}$.
$W_{k,0}$ can be computed with the method of \cite{CEO, EOFg} using the $x$-representation, while $W_{0,l}$  can be computed with the method of \cite{CEO, EOFg} using the $y$-representation, i.e. under the exchange
\beq
x\leftrightarrow y\quad .
\eeq

Therefore, in the following section, we improve the result of theorem \ref{thmixedtraces}, in order to prove that the diagrammatic rules of \cite{CEO, EOFg} are indeed symmetric under the exchange of $x$ and $y$.
In other words we prove theorem 7.1 of \cite{EOFg}, as announced in that article.


\section{Proof of the symmetry x-y of the algebraic invariants \texorpdfstring{$F^{(g)}(\curve)$}{Fg}}

Consider the two algebraic curves:
\beq
\hat\curve(x,y)=\curve(x,y)
\qquad  {\rm and} \qquad
\check\curve(x,y)=\curve(y,x)
\eeq
In \cite{EOFg}, for any curve $\curve$ an infinite sequence of invariants $F^{(g)}$ was defined.
Here we consider those invariants for the 2 curves $\hat\curve$ and $\check\curve$.

In this section we prove the following theorem (which was announced in \cite{EOFg}):
\bt Symmetry under the exchange $x\leftrightarrow y$:
\beq
\encadremath{
F^{(g)}(\hat\curve) = F^{(g)}(\check\curve)
}\eeq
where the functional $F^{(g)}(\curve)$ is defined for any curve $\curve(x,y)$ in \cite{EOFg}.
\et

\subsection{Preliminaries}

For the curve $\hat\curve(x,y)=0$, we have defined in \cite{EOFg} an infinite sequence of meromorphic forms:
\beq
\hat{W}^{(g)}_{k}(p_1,\dots,p_k)= \left.{W}^{(g)}_{k}(p_1,\dots,p_k)\right|_{\hat\curve}
\eeq
with poles only at the zeroes $\bfa=\{ a_i\}$ of $dx$, and some free energies
\beq
\hat{F}^{(g)} = {1\over 2-2g}\,\Res_{p\to \bfa} \Phi(p) \hat{W}^{(g)}_{1}(p)
\eeq
where $\Phi$ is any antiderivative of $ydx$, $d\Phi= y dx$ and ${\displaystyle \Res_{p \to \bfa}}$ stands for $\sum_i {\displaystyle \Res_{p \to a_i}}$.

And likewise, for the curve $\check\curve(x,y)=0$, we have defined an infinite sequence of meromorphic forms:
\beq
\check{W}^{(g)}_{k}(q_1,\dots,q_k)= \left.{W}^{(g)}_{k}(q_1,\dots,q_k)\right|_{\check\curve}
\eeq
with poles only at the zeroes $\mathbf{b}=\{ b_i\}$ of $dy$, and some free energies
\beq
\check{F}^{(g)} = {1\over 2-2g}\,\Res_{q\to \bfb} \Psi(q) \check{W}^{(g)}_{1}(q)
\eeq
where $d\Psi= x dy$.

\medskip

Our first step is to extend those forms into two families of multilinear meromorphic forms similar to those of section 2 (i.e. mimicking the mixed traces of matrix models):
\beq
\hat{W}^{(g)}_{k,l}(p_1,\dots,p_k|q_1,\dots,q_l)
\and
\check{W}^{(g)}_{k,l}(p_1,\dots,p_k|q_1,\dots,q_l)
\eeq
such that:
\beq
\hat{W}^{(g)}_{k,0}=\hat{W}^{(g)}_{k}
\virg
\check{W}^{(g)}_{0,l}=\check{W}^{(g)}_{l}.
\eeq

Our second step, is to prove that:
\beq
\hat{W}^{(g)}_{k,l}=
\check{W}^{(g)}_{k,l}.
\eeq

Our third step, is to prove that:
\beq
\hat{W}^{(g)}_{k+1,l}({\bf p_K},p|{\bf q_L})+\check{W}^{(g)}_{k,l+1}({\bf p_K}|p,{\bf q_L})
= d_p\left( {A^{(g)}_{k,l}(p;{\bf p_K}|{\bf q_L})\over dx(p)dy(p)} \right)
\eeq
where $A^{(g)}_{k,l}(p;{\bf p_K}|{\bf q_L})$ has poles of degree at most 2 at the poles of $ydx$, so that
in particular for $k=l=0$ we have:
\beq
\hat{W}^{(g)}_{1,0}(p)+\check{W}^{(g)}_{0,1}(p)
= d_p\left( {A^{(g)}_{0,0}(p)\over dx(p)dy(p)} \right)
\eeq
where $A^{(g)}_{0,0}$ has poles of degree at most 2 at the poles of $ydx$.

This last step is sufficent to prove that
\beq
\hat{F}^{(g)} = \check{F}^{(g)}.
\eeq

\subsection{Definitions of mixed correlators \texorpdfstring{$\hat{W}^{(g)}_{k,l}$}{Wg} and \texorpdfstring{$\check{W}^{(g)}_{k,l}$}{wg}}

We define the initial terms:
\beq
\hat{E}_{0,0}^{(0)}(x,y) = \check{E}_{0,0}^{(0)}(x,y)  = \curve(x,y),
\eeq
\beq
\hat{H}^{(0)}_{0,0}(p,q) =\check{H}^{(0)}_{0,0}(p,q) = {\curve(x(p),y(q))\over (x(p)-x(q))(y(p)-y(q))},
\eeq
\beq
\hat{W}^{(0)}_{1,0}(p) = \hat{W}^{(0)}_{0,1}(p)
=\check{W}^{(0)}_{1,0}(p) = \check{W}^{(0)}_{0,1}(p) = 0,
\eeq
\beq
\hat{W}^{(0)}_{2,0}(p,q) = \hat{W}^{(0)}_{0,2}(p,q) = - \hat{W}^{(0)}_{1,1}(p,q) = B(p,q),
\eeq
and
\beq
\check{W}^{(0)}_{2,0}(p,q) = \check{W}^{(0)}_{0,2}(p,q) = - \check{W}^{(0)}_{1,1}(p,q) = B(p,q).
\eeq

Let us define recursively the following quantities for any $g,k,l\geq 0$:
\beq\label{defJkl}
\begin{array}{l}
J_{k,l}^{(g)}(p,q;{\bf p_K}| {\bf q_L}) := \cr
 {\displaystyle \sum_{m_1,m_2=0}^k \sum_{n_1,n_2 = 0}^l \sum_{h,h'=1}^g} \hat{W}_{m_1+1,n_1}^{(h)}(p,{\bf p_{M_1}}|{\bf q_{N_1}}) \times \cr
\qquad \qquad \times \check{W}_{m_2,n_2+1}^{(h')}({\bf p_{M_2}}|{\bf q_{N_2}},q) H_{k-m_1-m_2,l-n_1-n_2}^{(g-h-h')}(p,q;{\bf p_{K/ \{M_1 \bigcup M_2\} }}|{\bf q_{L/ \{N_1 \bigcup N_2\} }} )\cr
+ {\displaystyle \sum_{h=1}^{g-1}} \left[ (x(p)-x(q)) \hat{W}_{1,0}^{(h)}(p) dy(q) + (y(q)-y(p)) \check{W}_{0,1}^{(h)}(q) dx(p) \right] H_{k,l}^{(g-h)}(p,q;{\bf p_K}| {\bf q_L}) \cr
+ {\displaystyle \sum_{m=0}^k \sum_{n=0;mn \neq kl}^{l} \sum_{h=0}^{g}}  H_{k-m,l-n}^{(g-h)}(p,q;{\bf p_{K/M}}| {\bf q_{L/N}}) \times\cr
 \;\;\; \times \left[ (x(p)-x(q)) \hat{W}_{m+1,n}^{(h)}(p,{\bf p_M}|{\bf q_N}) dy(q) + (y(q)-y(p)) \check{W}_{m,n+1}^{(h)}({\bf p_M}|{\bf q_N},q) dx(p) \right] \cr
 + (x(p)-x(q)) H_{k+1,l}^{(g-1)}(p,q;p,{\bf p_K}|{\bf q_L}) dy(q) + (y(q)-y(p)) H_{k,l+1}^{(g-1)}(p,q;{\bf p_K}|{\bf q_L},q) dx(p) \cr
 +{\displaystyle \sum_{m=0}^{k} \sum_{n=0}^l \sum_{h=0}^{g-1}} \left[ \hat{W}_{m+1,n}^{(h)}(p,{\bf p_M}|{\bf q_N})  H_{k-m,l-n+1}^{(g-h-1)}(p,q;{\bf p_{K/M}}| {\bf q_{L/N}},q) \right. \cr
 + {1\over 2} \left( \hat{W}_{m+1,n+1}^{(h)}(p,{\bf p_M}|{\bf q_N},q) + \check{W}_{m+1,n+1}^{(h)}(p,{\bf p_M}|{\bf q_N},q) \right) H_{k-m,l-n}^{(g-h-1)}(p,q;{\bf p_{K/M}}|{\bf q_{L/N}})  \cr
 \left. + \check{W}_{m,n+1}^{(h)}({\bf p_M}|{\bf q_N},q)  H_{k-m+1,l-n}^{(g-h-1)}(p,q;p,{\bf p_{K/M}}|{\bf q_{L/N}}) \right]  + H_{k+1,l+1}^{(g-2)}(p,q;p,{\bf p_K}|  {\bf q_L},q) \cr
\end{array}
\eeq
and
\beq\label{defcalJkl}
\begin{array}{l}
{\cal{J}}_{k,l}^{(g)}(p,q;{\bf p_K}| {\bf q_L}) :=
J_{k,l}^{(g)}(p,q;{\bf p_K}| {\bf q_L}) \cr
- {\displaystyle \sum_{\alpha=1}^{k}} d_{p_\alpha} \left\{ {dx(p) \over x(p)-x(p_{\alpha})} \left[
(x(p_\alpha)-x(q))dy(q)\, H_{k-1,l}^{(g)}(p_\alpha,q;{\bf p_{K-\{\alpha\}}}|{\bf q_L}) {\phantom {\displaystyle \sum_{i,j=0}^{<kl}} }\right. \right. \cr
+ {\displaystyle \sum_{h=1}^{g}} \check{W}_{0,1}^{(h)}(q) H_{k-1,l}^{(g-h)}(p_\alpha,q;{\bf p_{K-\{\alpha\}}}|{\bf q_L}) \cr
\left. \left. + {\displaystyle \sum_h \sum_{i,j=0}^{<kl}} H_{i-1,j}^{(g-h)}(p_\alpha,q;{\bf p_{I-\{\alpha\}}}|{\bf q_J}) \check{W}_{k-i,l-j+1}^{(h)}({\bf p_{K-I}}|{\bf q_{L-J}},q) \right] \right\} \cr
- {\displaystyle \sum_{\beta=1}^{l}} d_{q_\beta} \left\{ {dy(q) \over y(q)-y(q_\beta)} \left[
(y(q_\beta)-y(p))dx(p)\, H_{k,l-1}(p,q_\beta;{\bf p_{K}}|{\bf q_{L-\{\beta\}}}) {\phantom {\displaystyle \sum_{i,j=0}^{<kl}} }\right. \right. \cr
+ {\displaystyle \sum_{h=1}^{g} }\hat{W}_{1,0}^{(h)}(p) H_{k,l-1}^{(g-h)}(p,q_\beta;{\bf p_{K}}|{\bf q_{L-\{\beta\}}}) \cr
\left. \left. + {\displaystyle \sum_{i,j=0}^{<kl} \sum_h} H_{i,j-1}^{(g-h)}(p,q_\beta;{\bf p_{I}}|{\bf q_{J-\{\beta\}}}) \hat{W}_{k-i+1,l-j}^{h}(p,{\bf p_{K-I}}|{\bf q_{L-J}}) \right] \right\} \cr
+ {\displaystyle \sum_{\alpha=1}^{k} \sum_{\beta=1}^{l}} d_{p_\alpha} d_{q_\beta} \left\{ {dx(p) \over x(p)-x(p_{\alpha})}
{dy(q) \over y(q)-y(q_\beta)} H_{k-1,l-1}(p_\alpha,q_\beta; {\bf p_{K-\{\alpha\}}}| {\bf q_{L-\{\beta\}}}) \right\} \cr
-  {\displaystyle \sum_{\alpha=1}^{k}} d_{p_\alpha} \left( {H_{k-1,l+1}^{(g-1)}(p_\alpha,q;{\bf p_{K-\{\alpha\}}}|{\bf q_L},q)dx(p)  \over x(p)-x(p_{\alpha})} \right)
- {\displaystyle \sum_{\beta=1}^{l}} d_{q_\beta} \left( {H_{k+1,l-1}^{(g-1)}(p,q_\beta;p,{\bf p_K}|{\bf q_{L-\{\beta\}}})dy(q)  \over y(q)-y(q_\beta)} \right).  \cr
\end{array}
\eeq

\br
Those expressions are not as complicated as they look. They are inspired from section 2. In the matrix model case of section 2, those expressions contain nearly all the terms we would obtain from inserting loop equation \ref{loopeqtdUH} into loop equation \ref{loopeqEtdU}, or equivalently, from inserting loop equation \ref{loopeqUH} into loop equation \ref{loopeqEU}.
However, here we are not in a matrix model, and we don't assume any of the equations \ref{loopeqtdUH} to \ref{loopeqEU}, in fact we are going to prove them.
\er

\medskip

Now we define:
\bea\label{defhatWkl}
&& \hat{W}_{k+1,l}^{(g)}(p,{\bf p_{K}}|{\bf q_L}) \cr
&:=&
  \Res_{s \to \bfa, {\bf q_L}} dS_{s,o}(p) \left[
  {1\over d_1}\sum_{j=1}^{d_1}\, {{\cal{J}}^{(g)}_{k,l}(s,\tilde{s}^j;{\bf p_K}|{\bf q_L}) \over U_{0,0}^{(0)}(s,y(s)) dy(s)}
+ {1\over d_2}\sum_{i=1}^{d_2}\, {{\cal{J}}^{(g)}_{k,l}(s^i,s;{\bf p_K}|{\bf q_L}) \over \widetilde{U}_{0,0}^{(0)}(x(s),s) dx(s)}\right]
,\cr
\eea
\bea\label{defcheckWkl}
&& \hat{W}_{k,l+1}^{(g)}({\bf p_{K}}|{\bf q_L},q) \cr
&:=&
  \Res_{s \to \bfb, {\bf p_K}} dS_{s,o}(q) \left[
  {1\over d_1}\sum_{j=1}^{d_1}\, {{\cal{J}}^{(g)}_{k,l}(s,\tilde{s}^j;{\bf p_K}|{\bf q_L}) \over U_{0,0}^{(0)}(s,y(s)) dy(s)}
+ {1\over d_2}\sum_{i=1}^{d_2}\, {{\cal{J}}^{(g)}_{k,l}(s^i,s;{\bf p_K}|{\bf q_L}) \over \widetilde{U}_{0,0}^{(0)}(x(s),s) dx(s)}\right]
,\cr
\eea
\beq\label{defGkl}
\begin{array}{l}
 G_{k,l}^{(g)}(p,q;{\bf p_K}|{\bf q_L}) \cr
:= J_{k,l}^{(g)}(p,q;{\bf p_K}| {\bf q_L})+ H_{0,0}^{(0)}(p,q) \cr
 \left[(x(p)-x(q))\hat{W}_{k+1,l}^{(g)}(p,{\bf p_K}|{\bf q_L})dy(q) + (y(q)-y(p)) \check{W}_{k,l+1}^{(g)}({\bf p_K}|{\bf q_L},q) dx(p) \right], \cr
\end{array}
\eeq

\beq\label{defcalGkl}
\begin{array}{l}
 {\cal{G}}_{k,l}^{(g)}(p,q;{\bf p_K}| {\bf q_L}) \cr
:= {\cal{J}}_{k,l}^{(g)}(p,q;{\bf p_K}| {\bf q_L})+ H_{0,0}^{(0)}(p,q) \cr
 \left[(x(p)-x(q)) \hat{W}_{k+1,l}^{(g)}(p,{\bf p_K}|{\bf q_L})dy(q) + (y(q)-y(p)) \check{W}_{k,l+1}^{(g)}({\bf p_K}|{\bf q_L},q) dx(p) \right], \cr
\end{array}
\eeq

\beq\label{defhatHkl}
{\widehat{H}_{k,l}^{(g)}(p,q;{\bf p_K}|{\bf q_L}) \over \curve(x(p),y(q))}:=
\Res_{r \to q,p^i}{{\cal{G}}_{k,l}^{(g)}(p,r;{\bf p_K}|{\bf q_L}) \over (y(q)-y(p)) (y(q)-y(r)) (x(p)-x(r)) H_{0,0}^{(0)}(p,r) dx(p)},
\eeq

\beq\label{defcheckHkl}
{\check{H}_{k,l}^{(g)}(p,q;{\bf p_K}|{\bf q_L}) \over \curve(x(p),y(q))}:=
\Res_{r \to p,\tilde{q}^j}{{\cal{G}}_{k,l}^{(g)}(r,q;{\bf p_K}|{\bf q_L}) \over (x(p)-x(q)) (x(p)-x(r)) (y(q)-y(r)) H_{0,0}^{(0)}(r,q) dy(q)},
\eeq
and
\beq
{H}_{k,l}^{(g)} = {\hat{H}_{k,l}^{(g)}+\check{H}_{k,l}^{(g)}\over 2}
\eeq
(we prove below that  $\hat{H}_{k,l}^{(g)}=\check{H}_{k,l}^{(g)}={H}_{k,l}^{(g)}$) as well as

\beq\label{defhatEkl}
{\widehat{E}_{k,l}^{(g)}(p,q,{\bf p_K}|{\bf q_L}) \over \curve(x(p),y(q))}:=
\Res_{r \to p^i}{{\cal{G}}_{k,l}^{(g)}(p,r;{\bf p_K}|{\bf q_L}) \over (y(q)-y(p)) (y(q)-y(r)) (x(p)-x(r)) H_{0,0}^{(0)}(p,r) dx(p)},
\eeq

\beq\label{defcheckEkl}
{\check{E}_{k,l}^{(g)}(p,q,{\bf p_K}|{\bf q_L}) \over \curve(x(p),y(q))}:=
\Res_{r \to \tilde{q}^j}{{\cal{G}}_{k,l}^{(g)}(r,q;{\bf p_K}|{\bf q_L}) \over (x(p)-x(q)) (x(p)-x(r)) (y(q)-y(r)) H_{0,0}^{(0)}(r,q) dy(q)},
\eeq

\bea\label{deftdUkl}
 - \td{U}^{(g)}_{k,l}(p,q;{\bf p_K}|{\bf q_L})
&:=& (y(q)-y(p))H^{(g)}_{k,l}(p,q;{\bf p_K}|{\bf q_L}) \cr
&& + \sum_h\sum_{I,J} {\hat{W}^{(h)}_{i+1,j}(p,{\bf p_I}|{\bf q_J})  H^{(g-h)}_{k-i,l-j}(p,q;{\bf p_{K/I}}|{\bf q_{L/J}})\over dx(p)} \cr
&& +  {H^{(g-1)}_{k+1,l}(p,q;p,{\bf p_K}|{\bf q_L})\over dx(p)^2} \cr
&& - \sum_m d_{p_m}\,{H^{(g)}_{k-1,l}(p_m,q;{\bf p_{K/\{m\} }}|{\bf q_L})\over x(p)-x(p_m)} \cr
\eea
and


\bea\label{defUkl}
 - U^{(g)}_{k,l}(p,q;{\bf p_K}|{\bf q_L})
&:=& (x(p)-x(q))H^{(g)}_{k,l}(p,q;{\bf p_K}|{\bf q_L}) \cr
&& + \sum_h\sum_{I,J} {\check{W}^{(h)}_{i,j+1}({\bf p_I} |{\bf q_J},q)  H^{(g-h)}_{k-i,l-j}(p,q;{\bf p_{K/I}}|{\bf q_{L/J}})\over dy(q)} \cr
&& +  {H^{(g-1)}_{k,l+1}(p,q;{\bf p_K}|{\bf q_L},q)\over dy(q)^2} \cr
&& - \sum_n d_{q_n}\,{H^{(g)}_{k,l-1}(p,q_n;{\bf p_{K}}|{\bf q_{L/\{n\}}})\over y(q)-y(q_n)} .\cr
\eea

Those definitions form a triangular system of definitions, and each term is well defined in a unique recursive way.

\medskip
\br
Definitions eq.\ref{deftdUkl} and eq.\ref{defUkl} coincide with loop equation \ref{loopeqtdUH} and  \ref{loopeqUH} in the matrix model case,
i.e. when $\curve$ is the classical spectral curve of the 2 matrix model.
\er

\subsection{Theorems}

\bt\label{theoremHkg}
For $2g+k+l\geq 3$, one has the following properties:
\begin{itemize}

\item $\hat{W}_{k,l}^{(g)}({\bf p_K}|{\bf q_L})$ (resp. $\check{W}_{k,l}^{(g)}({\bf p_K}|{\bf q_L})$) has poles only when
$p_i \to \bfa,{\bf q_L}$ and $q_j \to \bfb, {\bf p_K}$;

\item in any of the $k+l$ variables, the $\acycle$-cycle integrals vanish: $\oint_{\acycle} \hat{W}^{(g)}_{k,l} = \oint_{\acycle} \check{W}^{(g)}_{k,l} = 0$;

\item $\widehat{H}_{k,l}^{(g)}(p,q;{\bf p_K}|{\bf q_L}) = \check{H}_{k,l}^{(g)}(p,q;{\bf p_K}|{\bf q_L})$ has poles only when
$p \to q, \bfa, {\bf q_L} $ and $q \to p, \bfb,{\bf p_K}$, and
\beq
\widehat{E}_{k,l}^{(g)}(x(p),q;{\bf p_K}|{\bf q_L}) = \check{E}_{k,l}^{(g)}(p,y(q)  ;{\bf p_K}|{\bf q_L}) := E^{(g)}(x(p),y(q);{\bf p_K}|{\bf q_L})\eeq
is a polynomial of degree $d_1-1$ in $x(p)$ and $d_2-1$ in $y(q)$;

\item ${U}_{k,l}^{(g)}(p,y(q);{\bf p_K}|{\bf q_L})$ (resp. $\widetilde{U}_{k,l}^{(g)}(x(p),q;{\bf p_K};{\bf q_L})$) is a
polynomial in $y(q)$ (resp. $x(p)$) of degree $d_2-1$ (resp. $d_1-1$).



\end{itemize}
\et

\proof{
Let us proceed by induction on $2g+k+l$. Suppose that the properties are satisfied for any
$g', k', l'$ such that $2g'+k'+l'<2g+k+l$.
Let us prove that they are true for $g,k,l$. In order to make the proof more readable, we split it into pieces. Nevertheless, {\bf for every step, the
global recursion hypothesis is needed}.

\vs

We need the following lemma:
\bl\label{lemmaftdfsymetry}
The quantity
\beq
f^{(g)}_{k,l}(s;{\bf p_{K}};{\bf q_L})
:= {{\cal{J}}^{(g)}_{k,l}(s,\tilde{s}^j;{\bf p_{K}};{\bf q_L}) \over U_{0,0}^{(0)}(s,y(s)) dy(s)}
\eeq
is independent of $j\neq 0$, it is a meromorphic one-form in the variable $s$, with poles at $s=\bfa,{\bf q_L}$, and it vanishes to order at least $\deg(ydx)-1$ near the poles of $ydx$.

Similarly, the quantity
\beq
\td{f}^{(g)}_{k,l}(s;{\bf p_{K}};{\bf q_L})
:=  {{\cal{J}}^{(g)}_{k,l}(s^i,s;{\bf p_{K}};{\bf q_L}) \over \widetilde{U}_{0,0}^{(0)}(x(s),s) dx(s)}.
\eeq
is independent of $i\neq 0$, it is a meromorphic one-form in the variable $s$, with poles at $s=\bfb, {\bf q_L}$,  and it vanishes to order at least $\deg(xdy)-1$ near the poles of $xdy$.

Moreover one has:
\beq
\oint_{\acycle} \Big( f^{(g)}_{k,l}(s;{\bf p_{K}};{\bf q_L}) +\td{f}^{(g)}_{k,l}(s;{\bf p_{K}};{\bf q_L})\Big) = 0,
\eeq
\beq
\oint_{\bcycle} \Big(f^{(g)}_{k,l}(s;{\bf p_{K}};{\bf q_L}) +\td{f}^{(g)}_{k,l}(s;{\bf p_{K}};{\bf q_L})\Big) = 0
\eeq
and:
\bea\label{lemmasymftdfcauchy}
&& f^{(g)}_{k,l}(s;{\bf p_{K}};{\bf q_L}) +\td{f}^{(g)}_{k,l}(s;{\bf p_{K}};{\bf q_L})  \cr
&=& \Res_{q\to \bfa,\bfb,{\bf p_{K}},{\bf q_L}}  dS_{q,o}(s)
\Big( f^{(g)}_{k,l}(q;{\bf p_{K}};{\bf q_L}) +\td{f}^{(g)}_{k,l}(q;{\bf p_{K}};{\bf q_L}) \Big).
\eea

\el

{\noindent\bf Proof of the lemma:}
{

First of all, One can remark that the definition of ${\cal{J}}_{k,l}^{(g)}$ involves only quantities whose properties
are known by the recursion hypothesis. One can note that it can be written under the following forms:
\beq\label{calJkltdUgen}
\begin{array}{l}
 J_{k,l}^{(g)}(p,q;{\bf p_K}; {\bf q_L}) \cr
:= - {\displaystyle \sum_{h=1}^{g-1} \sum_{m=0}^{k} \sum_{n=0}^{l}} \check{W}_{m,n+1}^{(h)}({\bf p_M};q,{\bf q_N}) \widetilde{U}_{k-m,l-n}^{g-h}(x(p),q;{\bf p_{K/M}},{\bf q_{L/N}}) dx(p) \cr
 - {\displaystyle \sum_{m=0}^{k} \sum_{n=0,(m,n)\neq(0,0)}^{l}} \check{W}_{m,n+1}^{(0)}({\bf p_M};q,{\bf q_N}) \widetilde{U}_{k-m,l-n}^{g}(x(p),q;{\bf p_{K/M}},{\bf q_{L/N}}) dx(p)\cr
 - {\displaystyle \sum_{m=0}^{k} \sum_{n=0,(m,n)\neq (k,l)}^{l}}\check{W}_{m,n+1}^{(g)}({\bf p_M};q,{\bf q_N}) \widetilde{U}_{k-m,l-n}^{0}(x(p),q;{\bf p_{K/M}},{\bf q_{L/N}}) dx(p)\cr
 - \widetilde{U}_{k,l+1}^{(g-1)}(x(p),q;{\bf p_K};q,{\bf q_L}) dx(p)\cr
 + {\displaystyle \sum_{h=1}^{g-1} \sum_{m=0}^{k} \sum_{n=0}^{l}} d_{p_\alpha} \left({\check{W}_{m,n+1}^{(h)}({\bf p_M};q,{\bf q_N}) H_{k-m-1,l-n}^{g-h}(p_\alpha,q;{\bf p_{K/M/\{\alpha\}}},{\bf q_{L/N}}) dx(p) \over x(p)-x(p_\alpha)} \right) \cr
 + {\displaystyle \sum_{m=0}^{k} \sum_{n=0,(m,n)\neq(0,0)}^{l}} d_{p_\alpha} \left({\check{W}_{m,n+1}^{(0)}({\bf p_M};q,{\bf q_N}) H_{k-m-1,l-n}^{g}(p_\alpha,q;{\bf p_{K/M/\{\alpha\}}},{\bf q_{L/N}}) dx(p) \over x(p)-x(p_\alpha)} \right) \cr
 + {\displaystyle \sum_{m=0}^{k} \sum_{n=0,(m,n)\neq (k,l)}^{l}} d_{p_\alpha} \left({\check{W}_{m,n+1}^{(g)}({\bf p_M};q,{\bf q_N}) H_{k-m-1,l-n}^{0}(p_\alpha,q;{\bf p_{K/M/\{\alpha\}}},{\bf q_{L/N}}) dx(p) \over x(p)-x(p_\alpha)} \right) \cr
 + d_{p_\alpha} \left({H_{k-1,l+1}^{(g-1)}(p_\alpha,q;{\bf p_{K/\{\alpha\}}};q,{\bf q_L}) dx(p) \over x(p)-x(p_\alpha)} \right)\cr
 + (x(p)-x(q)) \left[ {\displaystyle \sum_{h=1}^{g-1} \sum_{m=0}^{k} \sum_{n=0}^{l}} \hat{W}_{m+1,n}^{(h)}(p,{\bf p_M};{\bf q_N}) dy(q) H_{k-m,l-n}^{(g-h)}(p,q;{\bf p_{K/M}};{\bf q_{L/N}}) \right. \cr
 + {\sum_{m=0}^{k} \sum_{n=0,(m,n)\neq(k,l)}^{l}} \hat{W}_{m+1,n}^{(0)}(p,{\bf p_M};{\bf q_N}) dy(q) H_{k-m,l-n}^{(g)}(p,q;{\bf p_{K/M}};{\bf q_{L/N}}) \cr
 + {\sum_{m=0}^{k} \sum_{n=0,(m,n)\neq(0,0)}^{l}} \hat{W}_{m+1,n}^{(g)}(p,{\bf p_M};{\bf q_N}) dy(q) H_{k-m,l-n}^{(0)}(p,q;{\bf p_{K/M}};{\bf q_{L/N}}) \cr
 \left. + H_{k+1,l}^{(g-1)}(p,q;p,{\bf p_K};{\bf q_L})dy(q) \right] \cr
\end{array}
\eeq

and
\beq\label{calJklUgen}
\begin{array}{l}
 {\cal{J}}_{k,l}^{(g)}(p,q;{\bf p_K}; {\bf q_L}) \cr
= - {\displaystyle \sum_{h=1}^{g-1} \sum_{m=0}^{k} \sum_{n=0}^{l}} \check{W}_{m,n+1}^{(h)}({\bf p_M};q,{\bf q_N}) \widetilde{U}_{k-m,l-n}^{g-h}(x(p),q;{\bf p_{K/m}},{\bf q_{L/N}}) dx(p) \cr
 - {\displaystyle \sum_{m=0}^{k} \sum_{n=0,(m,n)\neq(0,0)}^{l}} \check{W}_{m,n+1}^{(0)}({\bf p_M};q,{\bf q_N}) \widetilde{U}_{k-m,l-n}^{g}(x(p),q;{\bf p_{K/m}},{\bf q_{L/N}}) dx(p)\cr
 - {\displaystyle \sum_{m=0}^{k} \sum_{n=0,(m,n)\neq (k,l)}^{l}}\check{W}_{m,n+1}^{(g)}({\bf p_M};q,{\bf q_N}) \widetilde{U}_{k-m,l-n}^{0}(x(p),q;{\bf p_{K/m}},{\bf q_{L/N}}) dx(p)\cr
 - \widetilde{U}_{k,l+1}^{(g-1)}(x(p),q;{\bf p_K};q,{\bf q_L}) dx(p)\cr
 + d_{q_\beta} \left( {\widetilde{U}_{k,l-1}^{(g)}(p,q_\beta;{\bf p_K};{\bf q_{L/\{\beta\}}}) \over y(q)-y(q_\beta)}dx(p)dy(q) \right) \cr
 - d_{p_\alpha} \left( {x(p_\alpha)-x(q) \over x(p)-x(p_\alpha)} H_{k-1,l}^{(g)}(p_\alpha,q;{\bf p_{K:\{\alpha\}}},{\bf q_L})dx(p)dy(q) \right) \cr
 + (x(p)-x(q)) \left[ {\displaystyle \sum_{h=1}^{g-1} \sum_{m=0}^{k} \sum_{n=0}^{l}} \hat{W}_{m+1,n}^{(h)}(p,{\bf p_M};{\bf q_N}) dy(q) H_{k-m,l-n}^{(g-h)}(p,q;{\bf p_{K/M}};{\bf q_{L/N}}) \right. \cr
 + {\displaystyle \sum_{m=0}^{k} \sum_{n=0,(m,n)\neq(k,l)}^{l}} \hat{W}_{m+1,n}^{(0)}(p,{\bf p_M};{\bf q_N}) dy(q) H_{k-m,l-n}^{(g)}(p,q;{\bf p_{K/M}};{\bf q_{L/N}}) \cr
 + {\displaystyle \sum_{m=0}^{k} \sum_{n=0,(m,n)\neq(0,0)}^{l}} \hat{W}_{m+1,n}^{(g)}(p,{\bf p_M};{\bf q_N}) dy(q) H_{k-m,l-n}^{(0)}(p,q;{\bf p_{K/M}};{\bf q_{L/N}}) \cr
 \left. + H_{k+1,l}^{(g-1)}(p,q;p,{\bf p_K};{\bf q_L}) \right]dy(q). \cr
\end{array}
\eeq

Thanks to the properties implied by the recursion hypothesis ($U$ and $\td{U}$ are polynomials), one has:
\bea\label{calJkltdU}
&& {\cal{J}}_{k,l}^{(g)}(q^i,q;{\bf p_K}; {\bf q_L}) \cr
&=& - \sum_{h=1}^{g-1} \sum_{m=0}^{k} \sum_{n=0}^{l} \check{W}_{m,n+1}^{(h)}({\bf p_M};q,{\bf q_N}) \widetilde{U}_{k-m,l-n}^{g-h}(x(q),q;{\bf p_{K/M}},{\bf q_{L/N}}) dx(q) \cr
&& - \sum_{m=0}^{k} \sum_{n=0,(m,n)\neq(0,0)}^{l} \check{W}_{m,n+1}^{(0)}({\bf p_M};q,{\bf q_N}) \widetilde{U}_{k-m,l-n}^{g}(x(q),q;{\bf p_{K/M}},{\bf q_{L/N}}) dx(q)\cr
&& - \sum_{m=0}^{k} \sum_{n=0,(m,n)\neq (k,l)}^{l}\check{W}_{m,n+1}^{(g)}({\bf p_M};q,{\bf q_N}) \widetilde{U}_{k-m,l-n}^{0}(x(q),q;{\bf p_{K/M}},{\bf q_{L/N}}) dx(q)\cr
&& - \widetilde{U}_{k,l+1}^{(g-1)}(x(q),q;{\bf p_K};q,{\bf q_L}) dx(q)\cr
&& + d_{q_\beta} \left( {\widetilde{U}_{k,l-1}^{(g)}(x(q),q_\beta;{\bf p_K};{\bf q_{L/\{\beta\}}}) \over y(q)-y(q_\beta)} \right)dx(q)dy(q) \cr
&& + d_{p_\alpha} \left( H_{k-1,l}^{(g)}(p_\alpha,q;{\bf p_{K/\{\alpha\}}},{\bf q_L}) \right)dx(q)dy(q) \cr
\eea
for any non vanishing $i$.
Thus this quantity does not depend on $i$, and $\td{f}$ is clearly a meromorphic 1-form, whose poles can be easily seen on this expression using the recursion hypothesis.

The same considerations give the equivalent through the exchange of $x \leftrightarrow y$:
\bea\label{calJklU}
&& {\cal{J}}_{k,l}^{(g)}(p,\td{p}^{j};{\bf p_K}; {\bf q_L}) \cr
&=& - \sum_{h=1}^{g-1} \sum_{m=0}^{k} \sum_{n=0}^{l} \hat{W}_{m+1,n}^{(h)}(p,{\bf p_M};{\bf q_N}) {U}_{k-m,l-n}^{g-h}(p,y(p);{\bf p_{K/M}},{\bf q_{L/N}}) dy(p) \cr
&& - \sum_{m=0}^{k} \sum_{n=0,(m,n)\neq(0,0)}^{l} \hat{W}_{m+1,n}^{(0)}(p,{\bf p_M};{\bf q_N}) {U}_{k-m,l-n}^{g}(p,y(p);{\bf p_{K/M}},{\bf q_{L/N}}) dy(p)\cr
&& - \sum_{m=0}^{k} \sum_{n=0,(m,n)\neq (k,l)}^{l}\hat{W}_{m+1,n}^{(g)}(p,{\bf p_M};{\bf q_N}) {U}_{k-m,l-n}^{0}(p,y(p);{\bf p_{K/M}},{\bf q_{L/N}}) dy(p)\cr
&& - U_{k+1,l}^{(g-1)}(p,y(p);p,{\bf p_K};{\bf q_L}) dy(p)\cr
&& + d_{p_\alpha} \left( {U_{k-1,l}^{(g)}(p_\alpha,y(p);{\bf p_{K/\{\alpha\}}};{\bf q_{L}}) \over x(p)-x(p_\alpha)} \right)dx(p)dy(p) \cr
&& + d_{q_\beta} \left( H_{k,l-1}^{(g)}(p,q_\beta;{\bf p_{K/\{\alpha\}}},{\bf q_{L/\{\beta\}}}) \right)dx(p)dy(p) \cr
\eea
This quantity does not depend on $j$, and $f$ is clearly a meromorphic 1-form, whose poles can be easily seen on this expression using the recursion hypothesis.

\bigskip

The fact that the $\acycle$ and $\bcycle$ cycle integrals vanish comes from the symmetry $x\leftrightarrow y$.
Indeed under the symmetry $x\leftrightarrow y$,  $f$ is changed to $\td{f}$ and $\td{f}$ is changed to $f$.
At the same time the $\acycle$-cycles are changed to $-\acycle$ because $2i\pi \epsilon = \oint_{\acycle} ydx = - \oint_{\acycle} xdy$, and  the $\bcycle$-cycles are changed to $-\bcycle$ in order to form a canonical basis.
Therefore, the $\acycle$ and $\bcycle$ cycle integrals of $f+\td{f}$ vanish.

\medskip
Equation \ref{lemmasymftdfcauchy} simply comes from Cauchy residue formula and Riemann's bilinear identity.

The fact that $f$ vanishes to order at least $\deg(ydx)-1$ near a pole $\alpha$ of $ydx$ follows from the definition of ${\cal J}$:
\bea
&&{{\cal J}_{k,l}^{(g)}(p,\td{p}^j;{\bf p_K}| {\bf q_L})\over dx(p) dy(p)} \sim_{p \to \alpha} \cr
&\sim_{p\to \alpha} & {x(p)-x(\td{p}^j)\over dx(p)} \,\Big(
 \sum_{m=0}^k \sum_{n=0}^{l} \sum_{h=0}^{g}  \;\;\;  \hat{W}_{m+1,n}^{(h)}(p,{\bf p_M}|{\bf q_N})  H_{k-m,l-n}^{(g-h)}(p,\td{p}^j;{\bf p_{K/M}}| {\bf q_{L/N}}) \cr
&& + H_{k+1,l}^{(g-1)}(p,\td{p}^j;p,{\bf p_K}|{\bf q_L}) \Big)\cr
&& - \sum_{\alpha=1}^{k} d_{p_\alpha} \left( {(x(p_\alpha)-x(\td{p}^j)) \over x(p)-x(p_{\alpha})}
\, H_{k-1,l}^{(g)}(p_\alpha,\td{p}^j;{\bf p_{K-\{\alpha\}}}|{\bf q_L})  \right) \cr
&& - \sum_{\beta=1}^{l} d_{q_\beta} \left( {(y(q_\beta)-y(p)) \over y(p)-y(q_\beta)}
\, H_{k,l-1}(p,q_\beta;{\bf p_{K}}|{\bf q_{L-\{\beta\}}})  \right) \cr
\eea
which is at most finite if $p$ approaches a pole $\alpha$ of $ydx$.
Then it implies that $f^{(g)}_{k,l}(p;{\bf p_{K}};{\bf q_L})
= {{\cal{J}}^{(g)}_{k,l}(p,\tilde{p}^j;{\bf p_{K}};{\bf q_L}) \over U_{0,0}^{(0)}(s,y(s)) dy(p)}$
vanishes at order at least $\deg(ydx)-1$.

The same holds for $\td{f}$.

\begin{flushright}
$\square$
\end{flushright}
}

\bigskip

$\bullet$ {\large \bf $W_{k,l}^{(g)}$ has poles only when
$p_i \to \bfa,{\bf q_L}$ and $q_j \to \bfb, {\bf p_K}$, and $\oint_{\acycle} W^{(g)}_{k,l} = 0$.}

From the definition eq.\ref{defhatWkl}, it is  clear that $\hat{W}_{k+1,l}^{(g)}(p,p_1,\dots,p_k|q_1,\dots,q_l)$ is finite when $p$ is not close to a branch point or to one of the $q_j$'s, and becomes infinite only if the integration contour is pinched. Thus in the variable $p$, the only poles of $\hat{W}_{k+1,l}^{(g)}(p,p_1,\dots,p_k|q_1,\dots,q_l)$ are at $p=\bfa, {\bf q_L}$.

The poles of $\hat{W}_{k+1,l}^{(g)}(p,p_1,\dots,p_k|q_1,\dots,q_l)$ in any other variable, follow from the recursion hypothesis, and thus they are at $p_i=\bfa, {\bf q_L}$, and at $q_j=\bfb, p, {\bf p_K}$.

The fact that $\oint_{\acycle} \hat{W}^{(g)}_{k+1,l} = 0$ when one integrates over the first variable comes from the fact that this is a property of $dS$, and in the other variables it comes from the recursion hypothesis.

By a symmetric argument, the same holds for $\check{W}_{k,l+1}^{(g)}(p_1,\dots,p_k|q_1,\dots,q_l,p)$, and we see that $\hat{W}_{k,l}^{(g)}$ and $\check{W}_{k,l}^{(g)}$ have the same poles.

We have (from the Cauchy residue formula and Riemann bilinear identity):
\beq\label{WhWcid}
\hat{W}_{k+1,l}^{(g)}(p,{\bf p_{K}}|{\bf q_L}) + \check{W}_{k,l+1}^{(g)}({\bf p_{K}}|{\bf q_L}, p) =
f^{(g)}_{k,l}(p;{\bf p_{K}}|{\bf q_L}) +
\td{f}^{(g)}_{k,l}(p;{\bf p_{K}}|{\bf q_L}).
\eeq

\vs

$\bullet$ {\bf \large $\widehat{H}_{k,l}^{(g)}(p,q;{\bf p_K}| {\bf q_L}) = \check{H}_{k,l}^{(g)}(p,q;{\bf p_K}| {\bf q_L})$.}

\vs

One has:
\beq
\begin{array}{l}
 {\widehat{H}_{k,l}^{(g)}(p,q;{\bf p_K}; {\bf q_L}) \over \curve(x(p),y(q)) } = \cr
= \Res_{r \to q,p^i} {{\cal{G}}_{k,l}^{(g)}(p,r;{\bf p_K};{\bf q_L}) \over (y(q)-y(p)) (y(q)-y(r)) (x(p)-x(r)) H_{0,0}^{(0)}(p,r) dx(p)} \cr
= \Res_{r \to q,p^i} \Res_{s \to p} {{\cal{G}}_{k,l}^{(g)}(s,r;{\bf p_K};{\bf q_L}) \over (y(q)-y(p)) (y(q)-y(r)) (x(s)-x(r)) (x(s)-x(p)) H_{0,0}^{(0)}(s,r) } \cr
= \Res_{r \to q,p^i} \Res_{s \to p, \tilde{q}^j} {{\cal{G}}_{k,l}^{(g)}(s,r;{\bf p_K};{\bf q_L}) \over (y(q)-y(p)) (y(q)-y(r)) (x(s)-x(r)) (x(s)-x(p)) H_{0,0}^{(0)}(s,r) } \cr
\end{array}\eeq
where the last equality holds because the integrant has no pole when $s \to \tilde{q}^j$.
Then
\beq
\begin{array}{l}
 {\widehat{H}_{k,l}^{(g)}(p,q;{\bf p_K}; {\bf q_L})\over \curve(x(p),y(q)) } = \cr
= \Res_{r \to q,p^i} \Res_{s \to p, \tilde{q}^j} {{\cal{G}}_{k,l}^{(g)}(s,r;{\bf p_K};{\bf q_L}) \over (y(q)-y(r)) (x(s)-x(p)) H_{0,0}^{(0)}(s,r) }
 \left[ {1 \over (y(r)-y(s)) (x(p)-x(q))} \right. \cr
 \;\;  \left. + { 1 \over (x(s)-x(r)) (y(q)-y(p))} - {1 \over (y(r)-y(s)) (x(p)-x(q))} \right]\cr
= \Res_{r \to q,p^i} \Res_{s \to p, \tilde{q}^j} {{\cal{G}}_{k,l}^{(g)}(s,r;{\bf p_K};{\bf q_L}) \over (y(q)-y(r)) (x(s)-x(p)) (y(r)-y(s)) (x(p)-x(q)) H_{0,0}^{(0)}(s,r) } \cr
 + \sum_{i=1}^{d_2} {{\cal{G}}_{k,l}^{(g)}(p,p^i;{\bf p_K};{\bf q_L}) \over (y(q)-y(p)) (y(q)-y(p^i)) H_{0,0}^{(0)}(p,p^i) dx(p)^2}.\cr
\end{array}\eeq
Note that the first term corresponds exactly to ${\check{H}_{k,l}^{(g)}(p,q;{\bf p_K}; {\bf q_L}) \over \curve(x(p),y(q)) }$
with the integration contours for $r$ and $s$ exchanged.
However, the poles of the integrand are known and thus:
\bea
\Res_{r \to q,p^i} \Res_{s \to p, \tilde{q}^j} &=& \Res_{r \to q} \Res_{s \to p} + \Res_{r \to p^i} \Res_{s \to\tilde{q}^j}
+ \Res_{r \to q} \Res_{s \to \tilde{q}^j} + \Res_{r \to p^i} \Res_{s \to p} \cr
&=& \Res_{s \to p} \Res_{r \to q} + \Res_{s \to\tilde{q}^j} \Res_{r \to p^i}
+ \sum_{j\neq 0} \Res_{\tilde{r}^j \to \tilde{q}^j} \Res_{s \to \tilde{q}^j} + \sum_{i \neq 0} \Res_{r^i \to p} \Res_{s \to p} \cr
&=& \Res_{s \to p} \Res_{r \to q} + \Res_{s \to\tilde{q}^j} \Res_{r \to p^i}
+ \sum_{j\neq 0}  \Res_{s \to \tilde{q}^j} \Res_{\tilde{r}^j \to \tilde{q}^j} + \sum_{j\neq 0}  \Res_{s \to \tilde{q}^j} \Res_{\tilde{r}^j \to s} \cr
&& \;\;  + \sum_{i \neq 0}  \Res_{s \to p} \Res_{r^i \to p} + \sum_{i \neq 0}  \Res_{s \to p} \Res_{r^i \to s} \cr
&=& \Res_{s \to p, \tilde{q}^j} \Res_{r \to q,p^i} + \sum_{j\neq 0}  \Res_{s \to \tilde{q}^j} \Res_{\tilde{r}^j \to s} + \sum_{i \neq 0}  \Res_{s \to p} \Res_{r^i \to s}. \cr
\eea
The last term does not contribute because the integrant is regular when $r^i \to s$, thus
\beq
\begin{array}{l}
 {\widehat{H}_{k,l}^{(g)}(p,q;{\bf p_K}; {\bf q_L}) \over \curve(x(p),y(q)) }  \cr
= {\check{H}_{k,l}^{(g)}(p,q;{\bf p_K}; {\bf q_L}) \over \curve(x(p),y(q)) }
+ \sum_{i=1}^{d_2} {{\cal{G}}_{k,l}^{(g)}(p,p^i;{\bf p_K};{\bf q_L}) \over (y(q)-y(p)) (y(q)-y(p^i)) H_{0,0}^{(0)}(p,p^i) dx(p)^2}\cr
 + \sum_{j\neq 0}  \Res_{s \to \tilde{q}^j} \Res_{\tilde{r}^j \to s} {{\cal{G}}_{k,l}^{(g)}(s,r;{\bf p_K};{\bf q_L}) \over (y(q)-y(r)) (x(s)-x(p)) (y(r)-y(s)) (x(p)-x(q)) H_{0,0}^{(0)}(s,r) } \cr
= {\check{H}^{(g)}(p,y(q)) \over \curve(x(p),y(q)) }
 + \sum_{i=1}^{d_2} {{\cal{G}}_{k,l}^{(g)}(p,p^i;{\bf p_K};{\bf q_L}) \over (y(q)-y(p)) (y(q)-y(p^i)) H_{0,0}^{(0)}(p,p^i) dx(p)^2} \cr
 + \sum_{j=1}^{d_1} {{\cal{G}}_{k,l}^{(g)}(\tilde{q}^j,q;{\bf p_K};{\bf q_L}) \over (x(p)-x(q)) (x(\tilde{q}^j)-x(p)) H_{0,0}^{(0)}(\tilde{q}^j,q) dy(q)^2}\cr
= {\check{H}^{(g)}(p,y(q)) \over \curve(x(p),y(q)) } \cr
 + \sum_{i=1}^{d_2} {(y(p)-y(p^i)) \over (y(q)-y(p)) (y(q)-y(p^i))}\,{\td{f}_{k,l}^{(g)}(p^i;{\bf p_K}| {\bf q_L})
 - \check{W}_{k,l+1}^{(g)}({\bf p_K}|{\bf q_L},p^i) \over  dx(p)} \cr
 + \sum_{j=1}^{d_1} {(x(q)-x(\td{q}^j)) \over (x(p)-x(q)) (x(p)-x(\td{q}^j))}\,{f_{k,l}^{(g)}(\td{q}^j|{\bf p_K}; {\bf q_L})
 - \hat{W}_{k+1,l}^{(g)}(\td{q}^j,{\bf p_K}|{\bf q_L}) \over  dy(q)} \cr
\end{array}\eeq

Notice from \eq{WhWcid}, that
\bea
g_{k,l}^{(g)}(s;{\bf p_K}| {\bf q_L})
&:=&\td{f}_{k,l}^{(g)}(s;{\bf p_K}|{\bf q_L})  - \check{W}_{k,l+1}^{(g)}({\bf p_K}|{\bf q_L},s)\cr
&=&
- f_{k,l}^{(g)}(s;{\bf p_K}| {\bf q_L}) + \hat{W}_{k+1,l}^{(g)}(s,{\bf p_K}|{\bf q_L})\cr
\eea
is a holomorphic 1-form in $s$, i.e. it has no poles.
We have:
\bea
&& {\widehat{H}_{k,l}^{(g)}(p,q;{\bf p_K}| {\bf q_L}) \over \curve(x(p),y(q)) } - {\check{H}_{k,l}^{(g)}(p,q;{\bf p_K}|{\bf q_L}) \over \curve(x(p),y(q)) } \cr
&=&  \sum_{i=1}^{d_2} {(y(p)-y(p^i)) \over (y(q)-y(p)) (y(q)-y(p^i))}\,{g_{k,l}^{(g)}(p^i;{\bf p_K}| {\bf q_L})
 \over  dx(p)} \cr
&& - \sum_{j=1}^{d_1} {(x(q)-x(\td{q}^j)) \over (x(p)-x(q)) (x(p)-x(\td{q}^j))}\,{g_{k,l}^{(g)}(\td{q}^j;{\bf p_K}| {\bf q_L})  \over  dy(q)} \cr
&=&  \sum_{i=1}^{d_2}  \Res_{s\to p^i} {(y(p)-y(s)) \over (y(q)-y(p)) (y(q)-y(s))}\,{g_{k,l}^{(g)}(s;{\bf p_K}| {\bf q_L})  \over  (x(s)-x(p))} \cr
&& - \sum_{j=1}^{d_1} \Res_{s\to \td{q}^j} {(x(q)-x(s)) \over (x(p)-x(q)) (x(p)-x(s))}\,{g_{k,l}^{(g)}(s;{\bf p_K}| {\bf q_L})  \over  (y(s)-y(q))} \cr
&=&  \sum_{i=0}^{d_2}  \Res_{s\to p^i}  \Big({(x(q)-x(s)) \over (x(p)-x(q)) }+{(y(p)-y(s)) \over (y(q)-y(p)) }\Big)\,{g_{k,l}^{(g)}(s;{\bf p_K}| {\bf q_L})  \over  (x(s)-x(p))(y(q)-y(s))} \cr
&=&  \sum_{i=0}^{d_2}  \Res_{s\to p^i}  \Big({1 \over (x(p)-x(q))(y(q)-y(s)) }-{1 \over (x(p)-x(s))(y(q)-y(s)) }\cr
&& +{1 \over (x(s)-x(p))(y(q)-y(p))} - {1 \over (x(s)-x(p))(y(q)-y(s))}\Big)\,{g_{k,l}^{(g)}(s;{\bf p_K}| {\bf q_L})  } \cr
&=&  \sum_{i=0}^{d_2}  \Res_{s\to p^i}  {1 \over (x(s)-x(p))(y(q)-y(p))} \,{g_{k,l}^{(g)}(s;{\bf p_K}| {\bf q_L})}  \cr
&=& 0
\eea

Therefore $\widehat{H}_{k,l}^{(g)}(p,q;{\bf p_K}| {\bf q_L}) = \check{H}_{k,l}^{(g)}(p,q;{\bf p_K}| {\bf q_L})={H}_{k,l}^{(g)}(p,q;{\bf p_K}| {\bf q_L})$.

\vs

$\bullet$ {\bf \large $\widehat{E}_{k,l}^{(g)}(p,q;{\bf p_K}| {\bf q_L}) = \check{E}_{k,l}^{(g)}(p,q;{\bf p_K}| {\bf q_L})$.}

\vs

We have from \eq{defhatEkl}
\beq\label{eqEklHG}
\widehat{E}_{k,l}^{(g)}(p,q,{\bf p_K}|{\bf q_L}) = (x(p)-x(q))(y(p)-y(q)) \widehat{H}_{k,l}^{(g)}(p,q,{\bf p_K}|{\bf q_L})   - {{\cal{G}}_{k,l}^{(g)}(p,q;{\bf p_K}|{\bf q_L}) \over  dx(p)dy(q)},
\eeq
and from \eq{defcheckEkl}:
\beq\label{eqEcheckklHG}
\check{E}_{k,l}^{(g)}(p,q,{\bf p_K}|{\bf q_L}) = (x(p)-x(q))(y(p)-y(q)) \check{H}_{k,l}^{(g)}(p,q,{\bf p_K}|{\bf q_L})   - {{\cal{G}}_{k,l}^{(g)}(p,q;{\bf p_K}|{\bf q_L}) \over  dx(p)dy(q)},
\eeq
so that $\widehat{E}_{k,l}^{(g)} = \check{E}_{k,l}^{(g)}$.

Moreover, one can see from \eq{defhatEkl} that $\widehat{E}_{k,l}^{(g)}(p,q;{\bf p_K}| {\bf q_L})$ is a polynomial of $y(q)$ while $\check{E}_{k,l}^{(g)}(p,q;{\bf p_K}| {\bf q_L})$ is a polynomial of $x(p)$, therefore
\beq
E_{k,l}^{(g)}(x(p),y(q);{\bf p_K}| {\bf q_L})
= \widehat{E}_{k,l}^{(g)}(p,q;{\bf p_K}| {\bf q_L}) = \check{E}_{k,l}^{(g)}(p,q;{\bf p_K}| {\bf q_L})
\eeq
is a polynomial in two variables.

\vs

$\bullet${\bf \large $U^{(g)}_{k,l}$ and $\td{U}^{(g)}_{k,l}$ are polynomials.}

\vs

\eq{eqEklHG}, \eq{eqEcheckklHG}, \eq{calJkltdUgen} and \eq{calJklUgen} imply that
\bea\label{loop1}
&& E^{(g)}_{k,l}(x(p),y(q);{\bf p_K}|{\bf q_L}) \cr
&=& (x(p)-x(q)) \td{U}^{(g)}_{k,l}(x(p),q;{\bf p_K}|{\bf q_L}) \cr
&& + \sum_h \sum_{I,J} {\check{W}^{(h)}_{i,j+1}({\bf p_I};{\bf q_J},q)  \td{U}^{(g-h)}_{k-i,l-j}(x(p),q;{\bf p_{K/I}}|{\bf q_{L/J}})\over dy(q)} \cr
&& + {\td{U}^{(g-1)}_{k,l+1}(x(p),q;{\bf p_K}|{\bf q_L},q)\over dy(q)} \cr
&& - \sum_m d_{q_m}\,{\td{U}^{(g)}_{k,l-1}(x(p),q_m;{\bf p_K}|{\bf q_{L/\{m\}}})\over y(q)-y(q_m)} \cr
&& - \sum_m d_{p_m}\,H^{(g)}_{k-1,l}(p_m,q;{\bf p_{K/\{m\} }}|{\bf q_L})
\eea
and
\bea\label{loop2}
&& E^{(g)}_{k,l}(x(p),y(q);{\bf p_K}|{\bf q_L}) \cr
&=& (y(q)-y(p)) U^{(g)}_{k,l}(p,y(q);{\bf p_K}|{\bf q_L}) \cr
&& + \sum_h \sum_{I,J} {\hat{W}^{(h)}_{i+1,j}(p,{\bf p_I};{\bf q_J})  U^{(g-h)}_{k-i,l-j}(p,y(q);{\bf p_{K/I}}|{\bf q_{L/J}})\over dx(p)} \cr
&& + {U^{(g-1)}_{k+1,l}(p,y(q);p,{\bf p_K}|{\bf q_L})\over dx(p)} \cr
&& - \sum_m d_{p_m}\,{U^{(g)}_{k-1,l}(p_m,y(q);{\bf p_{K/\{m\}}}|{\bf q_L}) \over x(p)-x(p_m)} \cr
&& - \sum_m d_{q_m}\,H^{(g)}_{k,l-1}(p,q_m;{\bf p_K}|{\bf q_{L/\{m\}}})
\eea
from which (together with the recursion hypothesis), we deduce that $U^{(g)}_{k,l}$ and $\td{U}^{(g)}_{k,l}$ are polynomials.

This proves the theorem \ref{theoremHkg}.
}

\bigskip

\bt\label{thsymWklg} 
Symmetry of the $W_{k,l}^{(g)}$.

For any $k,l,g$ we have:
\beq
\hat{W}_{k+1,l+1}^{(g-1)}(p,{\bf p_K}|{\bf q_L},q) = 
\check{W}_{k+1,l+1}^{(g-1)}(p,{\bf p_K}|{\bf q_L},q) 
\eeq
\et

\proof{
Let us prove it by recursion on $2g+k+l$.
Assume we have already proved it for any $g',k',l'$ such that $2g'+k'+l'<2g+k+l$.
\smallskip

Insert \eq{defUkl} into \eq{loop2} in order to eliminate the $U$'s, and then insert the result into \eq{eqEklHG}. Most of the terms cancel (in fact the definitions of $J_{k,l}^{(g)}$, ${\cal{J}}_{k,l}^{(g)}$, $G_{k,l}^{(g)}$ were designed for that purpose), and using the recursion hypothesis, the only term left is:
\beq
\check{W}_{k+1,l+1}^{(g-1)}(p,{\bf p_K}|{\bf q_L},q) = {1\over 2}\left( 
\check{W}_{k+1,l+1}^{(g-1)}(p,{\bf p_K}|{\bf q_L},q)+\hat{W}_{k+1,l+1}^{(g-1)}(p,{\bf p_K}|{\bf q_L},q)\right)
\eeq
which proves the theorem.
}

\bc\label{corWhatWcheck}
$\hat{W}_{k,l}^{(g)}({\bf p_K}|{\bf q_L}) =  \check{W}_{k,l}^{(g)}({\bf p_K}|{\bf q_L}) $ is a symmetric function of its variables $p_1,\dots, p_k$, and  a symmetric function of its variables $q_1,\dots, q_l$.
\ec

\proof{
It is clear from the definitions that $\check{W}_{k,l}^{(g)}({\bf p_K}|{\bf q_L}) $ is a symmetric function of its variables $p_1,\dots, p_k$, and that $\hat{W}_{k,l}^{(g)}({\bf p_K}|{\bf q_L})$ is a symmetric function of its variables $q_1,\dots, q_l$.
}

\bigskip

Now, we prove the following theorem:
\bt
\beq
\hat{W}_{k,0}^{(g)}({\bf p_K|})  =  \hat{W}_{k}^{(g)}({\bf p_K})
\eeq
and
\beq
\check{W}_{0,l}^{(g)}({\bf |q_L}) =  \check{W}_{l}^{(g)}({\bf q_L}).
\eeq
\et

\proof{
Write \eq{loop2} for {l=0}:
\bea
&& E^{(g)}_{k,0}(x(p),y(q);{\bf p_K}) \cr
&=& (y(q)-y(p)) U^{(g)}_{k,0}(p,y(q);{\bf p_K}) \cr
&& + \sum_h \sum_{I} {\hat{W}^{(h)}_{i+1,0}(p,{\bf p_I})  U^{(g-h)}_{k-i,0}(p,y(q);{\bf p_{K/I}})\over dx(p)} \cr
&& + {U^{(g-1)}_{k+1,0}(p,y(q);p,{\bf p_K})\over dx(p)}
 - \sum_m d_{p_m}\,{U^{(g)}_{k-1,0}(p_m,y(q);{\bf p_{K/\{m\}}}) \over x(p)-x(p_m)}. \cr
\eea
Using Lemma \ref{LemmaunicityEU}, we obtain:
\beq
\hat{W}_{k,0}^{(g)}({\bf p_K|})  =  \hat{W}_{k}^{(g)}({\bf p_K})
\eeq
The other equality is obtained by writing \eq{loop1} for $k=0$ and exchanging the roles of $x$ and $y$ in the Lemma \ref{LemmaunicityEU}.

}

\bt
\beq
\hat{W}_{k+1,l}^{(g)}(p,{\bf p_K|q_L})+\check{W}_{k,l+1}^{(g)}({\bf p_K|q_L},p)  =
d_p {A_{k,l}^{(g)}(p;{\bf p_K|q_L})\over dx(p) dy(p)}
\eeq
where $A_{k,l}^{(g)}(p;{\bf p_K|q_L})$ has at most simple poles when $p\to\bfalpha$.

\et

\proof{
From \eq{WhWcid}, it is easy to see that all contour integrals of
$\hat{W}_{k+1,l}^{(g)}(p,{\bf p_K|q_L})+\check{W}_{k,l+1}^{(g)}({\bf p_K|q_L},p) $ are vanishing, and thus it is the differential of some function.

The fact that $A_{k,l}^{(g)}(p;{\bf p_K|q_L})$ has at most simple poles when $p\to\bfalpha$, follows from
lemma \ref{lemmaftdfsymetry}.

}

\bt
\beq\label{resxyWgkl}
\Res_{p\to \bfalpha} x(p)y(p) \hat{W}^{(g)}_{k+1,l}(p,{\bf p_K}|{\bf q_L}) = 0,
\eeq
\beq\label{resxycheckWgkl}
\Res_{p\to \bfalpha} x(p)y(p) \hat{W}^{(g)}_{k,l+1}({\bf p_K}|{\bf q_L},p) = 0.
\eeq
\et

\proof{
By definition:
\beq
\hat{W}_{k+1,l}^{(g)}(p,{\bf p_{K}}|{\bf q_L})
=   \Res_{s \to \bfa, {\bf q_L}} dS_{s,o}(p) \, f_{k,l}^{(g)}(s;{\bf p_{K}}|{\bf q_L})
\eeq
and we have:
\bea
&& \Res_{p\to \bfalpha} x(p)y(p) \hat{W}^{(g)}_{k+1,l}(p,{\bf p_K}|{\bf q_L})   \cr
&=& \Res_{p\to \bfalpha}  \Res_{s \to \bfa, {\bf q_L}}  x(p)y(p) dS_{s,o}(p) \, f_{k,l}^{(g)}(s;{\bf p_{K}}|{\bf q_L})  \cr
&=& \Res_{s \to \bfa, {\bf q_L}} \Res_{p\to \bfalpha}   x(p)y(p) dS_{s,o}(p) \, f_{k,l}^{(g)}(s;{\bf p_{K}}|{\bf q_L})  \cr
&=& - \Res_{s \to \bfa, {\bf q_L}} (x(s)y(s)-x(o)y(o)) \, f_{k,l}^{(g)}(s;{\bf p_{K}}|{\bf q_L})  \cr
\eea
since $f_{k,l}^{(g)}$ vanishes near the poles of $ydx$ to order at least $\deg ydx-1$, the expression above has no other poles than $\bfa,{\bf q_L}$, and thus the total residue is zero.

}

\bt
For any $k$,$l$,$g$ such that $k+l+g \leq 1$, one has
\beq\begin{array}{rcl}
\Res_{p\to \bfa,{\bf q_L}} \Phi(p) \hat{W}^{(g)}_{k+1,l}(p,{\bf p_{K}}|{\bf q_L}) &=& \Res_{q\to \bfb,{\bf p_K}} \Psi(q) \check{W}^{(g)}_{k,l+1}({\bf p_{K}}|{\bf q_L},q)\cr
&=&(2-2g-k-l) \hat{W}_{k,l}^{(g)}({\bf p_{K}}|{\bf q_L}).\cr
\end{array}
\eeq
\et

\proof{

We have:
\bea
&& \Res_{p\to\bfa,{\bf q_L}} \Phi(p) \hat{W}^{(g)}_{k+1,l}(p,{\bf p_K}|{\bf q_L}) - \Res_{p\to\bfb,{\bf p_K}} \Psi(p) \check{W}^{(g)}_{k,l+1}({\bf p_K}|{\bf q_L},p) \cr
&=& \Res_{p\to\bfa,{\bf q_L}} x(p)y(p) \hat{W}^{(g)}_{k+1,l}(p,{\bf p_K}|{\bf q_L}) - \Res_{p\to\bfa,{\bf q_L}} \Psi(p) \hat{W}^{(g)}_{k+1,l}(p,{\bf p_K}|{\bf q_L}) \cr
&& \qquad - \Res_{p\to\bfb,{\bf p_K}} \Psi(p) \check{W}^{(g)}_{k,l+1}({\bf p_K}|{\bf q_L},p) \cr
&=& - \Res_{p\to\bfa,\bfb,{\bf p_K},{\bf q_L}} \Psi(p) (\hat{W}^{(g)}_{k+1,l}(p,{\bf p_K}|{\bf q_L}) + \check{W}^{(g)}_{k,l+1}({\bf p_K}|{\bf q_L},p)) \cr
&=&  \Res_{p\to\bfa,\bfb,{\bf p_K},{\bf q_L}} x(p)dy(p) {A^{(g)}_{k,l}(p;{\bf p_K}|{\bf q_L})\over dx(p)dy(p)}
 \cr
&=& - \Res_{p\to\bfalpha} x(p)dy(p) {A^{(g)}_{k,l}(p;{\bf p_K}|{\bf q_L})\over dx(p)dy(p)}
 \cr
&=&0.
\eea

The fact that $\Res_{p\to \bfa,{\bf q_L}} \Phi(p) \hat{W}^{(g)}_{k+1,l}(p,{\bf p_{K}}|{\bf q_L}) = (2-2g-k-l) \hat{W}_{k,l}^{(g)}({\bf p_{K}}|{\bf q_L})$, can be proved by recursion on $2g+k+l$ and using corolary \ref{corWhatWcheck}.

}

This allows to prove our main theorem:
\bt The $F^{(g)}$'s are symmetric under the exchange $x\leftrightarrow y$:
\beq\encadremath{
\hat{F}^{(g)} = \check{F}^{(g)}
}\eeq
\et
\proof{Indeed, we have:
\beq
(2-2g)\hat{F}^{(g)} = \Res_{\bfa} \Phi(p) \hat{W}^{(g)}_{1,0}(p)
\virg
(2-2g)\check{F}^{(g)} = \Res_{\bfb} \Psi(p) \check{W}^{(g)}_{0,1}(p).
\eeq

}

\subsection{Additional properties}

The following theorem  relates $H$ and $W$:
\bt
We have:
\beq
\hat{W}_{k+1,l}^{(g)}(p,{\bf p_{K}}|{\bf q_L})
= \Res_{q\to\bfalpha} {H_{k,l}^{(g)}(p,q;{\bf p_{K}}|{\bf q_L})\over H_{0,0}^{(0)}(p,q)} dy(q)
\eeq
\beq
\check{W}_{k,l+1}^{(g)}({\bf p_{K}}|{\bf q_L},q)
= \Res_{p\to\bfalpha} {H_{k,l}^{(g)}(p,q;{\bf p_{K}}|{\bf q_L})\over H_{0,0}^{(0)}(p,q)} dx(p).
\eeq
\et

\proof{
Multiply equation \ref{defUkl} by $dx(p)dy(q)/(y(q)-y(p))H_{0,0}^{(0)}(p,q)$ and take the residues at $q\to\bfalpha$.

}

\br
This theorem was expected from the matrix model property that
\beq
\tr {1 \over x-M_1} {1 \over y-M_2} \to {1 \over x} \tr {1 \over y-M_2}
\eeq
when $x \to \infty$.
\er

\section{Conclusion}

In this article, we have proved the $x\leftrightarrow y$ symmetry which was announced in \cite{EOFg}.
This symmetry has many applications, for instance in \cite{EOFg} it was used to recover the $(p,q)\leftrightarrow (q,p)$ duality of minimal models \cite{KharMar}, or to give a very short proof that Kontsevitch integral indeed depends only on odd times and satisfies KdV hierarchy \cite{IZK}.

In addition we have shown how to compute some family of mixed correlation functions of the 2-matrix model.

This could open the route to some matrix model approach to the understanding of boundary conformal field theory in higher genus.
In a forthcoming article, we shall introduce a similar algebraic geometry method to compute all possible mixed correlation functions \cite{EOallmixed}.

\bigskip

This work also raises many questions, and calls the following prospects:

$\bullet$ It would be interesting to see what the $H_{k,l}$ and $W_{k,l}$ correspond to for other matrix models (e.g. Kontsevitch's integral, chain of matrices), although we may guess that they also correspond to mixed traces expectation values in those cases.

$\bullet$ More interesting would be to understand what the $H_{k,l}^{(g)}$ and $W_{k,l}^{(g)}$ compute in algebraic geometry. Those should correspond to ``volume'' or ``intersection numbers of some moduli spaces'' ?

\subsection*{Acknowledgements}

We would like to thank Michel Berg\`ere and Aleix Prats Ferrer for fruitful discussions on this subject.
This work is partly supported by the Enigma European network MRT-CT-2004-5652, by the ANR project G\'eom\'etrie et int\'egrabilit\'e en physique math\'ematique ANR-05-BLAN-0029-01,
 by the Enrage European network MRTN-CT-2004-005616,
 by the European Science foundation through the Misgam program,
 by the French and Japaneese governments through PAI Sakura, by the Quebec government with the FQRNT.

\vfill\eject

\setcounter{section}{0}

\appendix{Spectral curve}

We recall that the curve $\curve(x,y)$, called the classical spectral curve, is given by a polynomial of the form:
\beq
\curve(x,y) = \sum_{j=0}^{d_2+1} \curve_{j}(x) y^{j}
\eeq
We define the ``quantum spectral curve'' as the formal power series:
\bea
\curve_N(x,y) = \sum_g N^{-2g}\,\curve^{(g)}(x,y)
\eea
where
\bea
\curve^{(g)}(x,y) = \curve_{d_2+1}(x) \sum_{r=1}^{d_2} \sum_{J_1\cup \dots \cup J_r= K} \sum_{g_1,\dots,g_r}\, \delta_{\sum_l (g_l+|J_l|-1),g\,}\, \prod_{l=1}^r\, {\td{W}_{|J_l|}}^{(g_l)}(p^{J_l})
\eea
with
\beq
K=\{1,\dots,d_2\}
\eeq
and
\beq
{\td{W}_{k}}^{(g)}(p_K) := W_{k}^{(g)}(p_K) + \delta_{k,1}\delta_{g,0} (y-Y(p_1))
\eeq
where $W_{k}^{(g)}(p_K)$ is the meromorphic form defined in \cite{EOFg} for the curve $\curve(x,y)$.

\bl\label{lemmaspectralcurve}
{For any $g$, $\curve^{(g)}(x,y)$ is a polynomial in $x$ and $y$, whose degrees are at most those of $\curve$.}
\el

\proof{
It is clear that $\curve_N(x,y)$ is a polynomial in $y$, and a rational function of $x$.
Let us prove that $\curve^{(g)}(x,y)$ is indeed a polynomial in $x$ for $g\geq 1$.
The coefficient of $y^k$ in $\curve^{(g)}(x,y)$ is:
\beq
\begin{array}{l}
{\curve_k^{(g)}(x) \over  \curve_{d_2+1}(x)} \cr
{\displaystyle = \sum_{J_0\subset K, |J_0|=k}\,\prod_{j\in J_0} y(p^j)\,\,\sum_{r=1}^{d_2-k} \sum_{J_1\cup \dots \cup J_r= K/J_0} \sum_{g_1,\dots,g_r}\, \delta_{\sum_l (g_l+|J_l|-1),g\,}\,  \prod_{l=1}^r\, {W_{|J_l|}}^{(g_l)}(p^{J_l})}\cr
\end{array}\eeq
First, notice that the product of $W$'s can have poles only at branch-points, and the product of $y$'s can have poles only at poles of $y$.
The poles of $y$ which are not poles of $x$, are killed by the prefactor $\curve_{d_2+1}(x)$, as they are in the classical curve $\curve(x,y)$.
Let us consider the poles at a branch-point $a$.
The only terms which might diverge at $p\to a$ are of either of the following forms
\begin{itemize}
\item
$(W_{1+|J|}^{(h)}(p,p^{J})+W^{(h)}_{1+|J|}(\bar{p},p^{J})) \times ({\rm reg})
$
where reg means a term with no poles at $p\to a$. This term is regular because of theorem 4.4 in \cite{EOFg}.

\item
or
$
(W^{(g_1)}_{1+|J_1|}(p,p^{J_1})W^{(h-g_1)}_{1+|J|-|J_1|}(\bar{p},p^{J/J_1})+W^{(h-1)}_{2+|J|}(p,\bar{p},p^{J})) \times ({\rm reg})
$
again, this expression is regular when $p\to a$, because of theorems 4.4 and 4.5 in \cite{EOFg}.

\end{itemize}

Thus, we have  proved that $\curve_k^{(g)}(x)$ is a rational function of $x$ whose only poles are the poles of $x$, i.e. it is a polynomial in $x$.

Consider a pole $\infty_x$ of $x$, the behavior of $\curve^{(g)}(x(p),y(p))$ when $p\to\infty_x$ is at most that of $\sum_{J_0\subset K}\,\prod_{j\in J_0} y(p^j)$.
Notice that $J_0$ cannot be equal to $K$ itself, because the product of the corresponding $W$'s vanishes (it contains no term), and $|J_0|$ cannot be equal to $|K|-1$, because the prefactor vanishes due to theorem 4.4 in \cite{EOFg}.
Thus, $|J_0|\leq |K|-2$, which implies that $\curve^{(g)}(x(p),y(p))dx(p)$ has a pole of degree at most that of $\curve_y(x(p),y(p))$, i.e. $\curve^{(g)}(x(p),y(p))$ is contained in the Newton's polytope of $\curve(x,y)$.
This means that
\beq
{\curve^{(g)}(x(p),y(p))\over \curve_y(x(p),y(p))}dx(p)
\eeq
is a holomorphic differential.

}

\appendix{Lemma: unicity of the solution of loop equations}

\bl\label{LemmaunicityEU}
The system of equations:
\bea\label{loop2lemma}
&& E^{(g)}_{k}(x(p),y(q);{\bf p_K}) \cr
&=& (y(q)-y(p)) U^{(g)}_{k}(p,y(q);{\bf p_K}) \cr
&& + \sum_h \sum_{I} {W^{(h)}_{i+1}(p,{\bf p_I})  U^{(g-h)}_{k-i}(p,y(q);{\bf p_{K/I}})\over dx(p)} \cr
&& + {U^{(g-1)}_{k+1}(p,y(q);p,{\bf p_K})\over dx(p)}
 - \sum_m d_{p_m}\,{U^{(g)}_{k-1}(p_m,y(q);{\bf p_{K/\{m\}}}) \over x(p)-x(p_m)} \cr
\eea
where:
\begin{itemize}
\item if $2g+k>2$, $W^{(g)}_{k+1}(p,{\bf p_K})$ has poles only at branchpoints in any of its variables, and vanishing $\acycle$-cycle integrals,
\item $E^{(g)}_{k}(x(p),y(q);{\bf p_K})$ is a polynomial in $x(p)$ of degree at most $d_1-1$, and a polynomials in $y(q)$ of degree at most $d_2-1$,
\item $U^{(g)}_{k}(p,y(q);{\bf p_K})$ is a polynomials in $y(q)$ of degree at most $d_2-1$,
\end{itemize}
 has a unique solution.

This solution is such that
\beq
W_{k}^{(g)}({\bf p_K})  =  \hat{W}_{k}^{(g)}({\bf p_K}).
\eeq

\el

{\bf Proof of the Lemma:}

{
{\bf Unicity:}

We prove it by recursion on $2g+k$. Assume it is already proved for any $g',k'$ such that $2g'+k<2g+k$.

At $p=q$, \eq{loop2lemma} gives:
\bea\label{Wgkloop2}
 W^{(g)}_{k+1}(p,{\bf p_K})
&=& {E^{(g)}_{k}(x(p),y(p);{\bf p_K}) dx(p) \over U^{(0)}_{0}(p,y(p))} \cr
&& - \sum_h \sum_{I} {W^{(h)}_{i+1}(p,{\bf p_I})  U^{(g-h)}_{k-i}(p,y(p);{\bf p_{K/I}})\over U^{(0)}_{0}(p,y(p))} \cr
&& - {U^{(g-1)}_{k+1}(p,y(p);p,{\bf p_K})\over U^{(0)}_{0}(p,y(p))}
 + \sum_m d_{p_m}\,{U^{(g)}_{k-1}(p_m,y(p);{\bf p_{K/\{m\}}})dx(p) \over (x(p)-x(p_m))U^{(0)}_{0}(p,y(p))} .\cr
\eea
Then write Cauchy residue formula:
\beq
W^{(g)}_{k+1}(p,{\bf p_K})  = - \Res_{r\to p}  dS_{r,o}(p)\,\,W^{(g)}_{k+1}(r,{\bf p_K}).
\eeq
Since we know the poles of $W^{(g)}_{k+1}(p,{\bf p_K}) $ and its $\acycle$-cycle integrals, we may move the integration contour using Riemann's bilinear identity and get:
\beq
W^{(g)}_{k+1}(p,{\bf p_K})  =  \Res_{r\to \bfa}  dS_{r,o}(p)\,\,W^{(g)}_{k+1}(r,{\bf p_K}).
\eeq
Now, we replace $W^{(g)}_{k+1}(r,{\bf p_K})$ by its value in \eq{Wgkloop2}.
We see that the term ${E^{(g)}_{k}(x(r),y(r);{\bf p_K}) dx(r) \over U^{(0)}_{0}(r,y(r))}$ has no pole at the branchpoints and does not contribute to the residue, and similarly the las term of \eq{Wgkloop2} does not contribute to the residue. We get:
\bea
W^{(g)}_{k+1}(p,{\bf p_K})
&=& - \Res_{r\to \bfa}  {dS_{r,o}(p)\over U^{(0)}_{0}(p,y(p))}
\,\,\Big( U^{(g-1)}_{k+1}(r,y(r);p,{\bf p_K}) \cr
&& + \sum_h \sum_{I}  W^{(h)}_{i+1}(r,{\bf p_I})  U^{(g-h)}_{k-i}(r,y(r);{\bf p_{K/I}}) \Big).
\eea
Since all the terms in the RHS are already known from the recursion hypothesis, this determines
$W^{(g)}_{k+1}(p,{\bf p_K})$ uniquely.
Then, we write \eq{loop2lemma} for $p=\td{q}^j$ with $j=1,\dots,d_1$:
\bea
&& E^{(g)}_{k}(x(\td{q}^j),y(q);{\bf p_K}) \cr
&=&  \sum_h \sum_{I} {W^{(h)}_{i+1}(\td{q}^j,{\bf p_I})  U^{(g-h)}_{k-i}(\td{q}^j,y(q);{\bf p_{K/I}})\over dx(\td{q}^j)} \cr
&& + {U^{(g-1)}_{k+1}(\td{q}^j,y(q);\td{q}^j,{\bf p_K})\over dx(\td{q}^j)}
 - \sum_m d_{p_m}\,{U^{(g)}_{k-1}(p_m,y(q);{\bf p_{K/\{m\}}}) \over x(\td{q}^j)-x(p_m)} \cr
\eea
since all terms in the RHS are uniquely determined, so is the LHS. And since
$E^{(g)}_{k}(x(p),y(q);{\bf p_K})$ is a polynomial in $x(p)$ of degree $d_1-1$ and we know its value in $d_1$ points, then $E^{(g)}_{k}(x(p),y(q);{\bf p_K})$ is uniquely determined.

Then, using \eq{loop2lemma} once again, we uniquely determine $U^{(g)}_{k}(p,y(q);{\bf p_K})$.

This proves the unicity for $g$ and $k$.

\medskip
{\bf Existence:}

Start from the meromorphic form 
$W_{k}^{(g)}(p_K)$  defined in \cite{EOFg} for the curve $\curve(x,y)$, and define:
\beq
{\td{W}_{k}}^{(g)}(p_K) := W_{k}^{(g)}(p_K)/dx(p_K) + \delta_{k,1}\delta_{g,0} (y-y(p_1))
\eeq

Then, let $K_0=\{0,1,\dots,d_2\}\cup K$ and $K_1=\{1,\dots,d_2\}\cup K$, and define:
\beq
\curve_k^{(g)}(x(p^0),y;p_K) 
= \curve_{d_2+1}(x) \sum_{r=1}^{d_2+1+k} \sum_{J_1\cup \dots \cup J_r= K_0} \sum_{g_1,\dots,g_r}\, \delta_{\sum_l (g_l+|J_l|-1),g\,}\, \prod_{l=1}^r\, {\td{W}_{|J_l|}}^{(g_l)}(p^{J_l})
\eeq
and:
\beq
U_k^{(g)}(p^0,y;p_K) 
= \curve_{d_2+1}(x) \sum_{r=1}^{d_2+k} \sum_{J_1\cup \dots \cup J_r= K_1} \sum_{g_1,\dots,g_r}\, \delta_{\sum_l (g_l+|J_l|-1),g\,}\, \prod_{l=1}^r\, {\td{W}_{|J_l|}}^{(g_l)}(p^{J_l}).
\eeq
It is clear that both $\curve_k^{(g)}(x,y;p_K)$ and $U_k^{(g)}(p,y;p_K)$ are polynomials in $y$ of degree at most $d_2-1$.
Following the same line as in lemma \ref{lemmaspectralcurve}, it is easy to get that $\curve_k^{(g)}(x,y;p_K)$ is also a polynomial in $x$ of degree at most $d_1-1$.

Therefore, the functions $\curve_k^{(g)}(x,y;p_K)$, $U_k^{(g)}(p,y;p_K)$ and $W_k^{(g)}(p_K)$ obey the requirements of lemma \ref{LemmaunicityEU}, and eq.\ref{loop2lemma} is clearly satisfied from the definitions of $\curve_k^{(g)}(x,y;p_K)$ and $U_k^{(g)}(p,y;p_K)$.
Thus, we have found an explicit solution of the system of lemma \ref{LemmaunicityEU}, which proves the existence.

}



\begin{thebibliography}{99}









\bibitem{ACKM}
J.Ambj{\o}rn, L.Chekhov, C.F.Kristjansen and Yu.Makeenko,
``Matrix model calculations beyond the spherical limit'',
{\em Nucl.Phys.} {\bf B404} (1993) 127--172; Erratum ibid. {\bf B449} (1995) 681,
hep-th/9302014.



\bibitem{BergereEyn} M. Berg\`ere and B. Eynard, ``Mixed correlation function and spectral curve for the 2-matrix model'',
{\emph J. Phys. A: Math. Gen.} {\bf 39} No 49 (8 December 2006) 15091-15134, math-ph/0605010.


\bibitem{Marco2} M. Bertola, ''Two-matrix model with semiclassical potentials and extended Whitham hierarchy'',
{\em  J.Phys.} {\bf A39} 8823-8856  (2006),  hep-th/0511295.


\bibitem{BEmixed} M. Bertola, B. Eynard, `` Mixed Correlation functions of the 2-Matrix Model'',
{\emph J. Phys.} {\bf A36} (2003) 7733-7750, hep-th/0303161.

\bibitem{MarcoF} M. Bertola, ''Free Energy of the Two-Matrix
Model/dToda Tau-Function'',
preprint CRM-2921 (2003), hep-th/0306184.


\bibitem{BIPZ} E. Brezin, C. Itzykson, G. Parisi, and J. Zuber,
{\em Comm. Math. Phys.} {\bf 59}, 35  (1978).



\bibitem{ec1loopF} L.Chekhov, B.Eynard,
``Hermitian matrix model free energy: Feynman graph technique for all genera'',
{\em J. High Energy Phys.} {\bf JHEP03} (2006) 014, hep-th/0504116.


\bibitem{CEO} L.Chekhov, B.Eynard and N.Orantin,
``Free energy topological expansion for the 2-matrix model'',
{\em J. High Energy Phys.} {\bf JHEP12} (2006) 053, math-ph/0603003.

\bibitem{DKK} J.M.Daul, V.Kazakov, I.Kostov,
``Rational Theories of 2D Gravity from the Two-Matrix Model'',
{\em Nucl.Phys.} {\bf B409} (1993) 311-338, hep-th/9303093.




\bibitem{Virasoro} F. David,
``Loop equations and nonperturbative effects in two-dimen\-sional quantum gravity''.
{\em Mod.Phys.Lett.} {\bf A5} (1990) 1019.

\bibitem{davidRMT} F. David, "Planar diagrams, two-dimensional lattice gravity and surface models",
Nuclear Physics B, Volume 257, p. 45-58.

\bibitem{ZJDFG} P. Di Francesco, P. Ginsparg, J. Zinn-Justin,
``2D Gravity and Random Matrices'',
{\em Phys. Rep.} {\bf 254}, 1 (1995).








\bibitem{eynloop1mat} B. Eynard, ``Topological expansion for the 1-hermitian matrix model correlation functions'',
JHEP/024A/0904, hep-th/0407261.

\bibitem{eynm2m} B. Eynard, ``Large N expansion of the 2-matrix model'',
{\em JHEP} {\bf 01} (2003) 051, hep-th/0210047.

\bibitem{eynm2mg1} B. Eynard, ``Large N expansion of the 2-matrix model,
multicut case'',
preprint SPHT03/106, ccsd-00000521, math-ph/0307052.

\bibitem{eylooprat} B.Eynard,
``Loop equations for the semiclassical 2-matrix model with hard edges'',
{\em J.Stat.Mech.} {\bf 0510} (2005) P006, math-ph/0504002.


\bibitem{eynprats} B. Eynard, A. Prats Ferrer, ``2-matrix versus complex matrix model, integrals over the unitary group as triangular integrals'',
{\emph Commun.Math.Phys.} {\bf 264} (2006) 115-144, hep-th/0502041.

\bibitem{eynchain} B. Eynard, ``Master loop equations, free energy and correlations for the chain of matrices'',
{\em J. High Energy Phys.}  {\bf JHEP11}(2003)018, hep-th/0309036.


\bibitem{eyno} B.Eynard, N.Orantin,
``Topological expansion of the 2-matrix model correlation functions: diagrammatic rules for a residue formula'',
{\em J. High Energy Phys.} {\bf JHEP12}(2005)034, math-ph/0504058.

\bibitem{eynform} B.Eynard,
`` Formal matrix integrals and combinatorics of maps'', \\
math-ph/0611087.


\bibitem{EOFg} B.Eynard, N.Orantin,
``Invariants of algebraic curves and topological expansion'', math-ph/0702045.

\bibitem{EObethe} B. Eynard, N. Orantin, ``Mixed correlation functions in the 2-matrix model, and the Bethe ansatz'',
{\emph JHEP} {\bf 0508} (2005) 028, hep-th/0504029.

\bibitem{EOallmixed} B. Eyner, N. Orantin, ``Whole topological expnasion of any correlation function in the two matrix model'', in preparation.

\bibitem{Farkas} H.M. Farkas, I. Kra, ''Riemann surfaces'' 2nd edition, Springer Verlag, 1992.

\bibitem{Fay} J.D. Fay, ''Theta functions on Riemann surfaces'', Springer Verlag, 1973.

\bibitem{IZK} C.Itzykson, J.B.Zuber,
``Combinatorics of the Modular Group II: The Kontsevich integrals'',
{\em Int.J.Mod.Phys.} {\bf A7} (1992) 5661-5705, hep-th/9201001 .



\bibitem{KazakovRMT} V.A. Kazakov, "Bilocal regularization of models of random surfaces"
Physics Letters B, Volume 150, Issue 4, p. 282-284.


\bibitem{KazakovIsing} V.A. Kazakov, ``Ising model on a dynamical
planar random lattice: exact solution'',
{\em Phys Lett.} {\bf A119}, 140-144 (1986).

\bibitem{KazMar} V.A. Kazakov, A. Marshakov, ''Complex Curve of the
Two Matrix Model and its Tau-function'',
{\em J.Phys.} {\bf A36} (2003) 3107-3136, hep-th/0211236.

\bibitem{KharMar} S.Kharchev, A.Marshakov, ``On $p-q$ duality and explicit solutions in $c<1$ 2d gravity models'',
hep-th/9303100.



\bibitem{Kri} I.Krichever
``The $\tau$-function of the universal Whitham
hierarchy, matrix models and topological field theories'',
{\em Commun.Pure Appl.Math.} {\bf47} (1992) 437; hep-th/9205110



\bibitem{staudacher} M. Staudacher,
`` Combinatorial solution of the 2-matrix model'',
{\em Phys. Lett.} {\bf B305} (1993) 332-338.


\bibitem{thooft} G. 't Hooft, {\em Nuc. Phys.} {\bf B72}, 461 (1974).


\bibitem{tutte} W.T. Tutte, ``A census of planar triangulations'',
{\em Can. J. Math.} {\bf 14} (1962) 21-38.

\bibitem{tutte2} W.T. Tutte, ``A census of planar maps'', 
{\em Can. J. Math.} {\bf 15} (1963) 249-271.




















\end{thebibliography}
\end{document}